\documentclass[prl,twocolumn]{revtex4}

\usepackage{amsmath}
\usepackage{amssymb}
\usepackage{amsfonts}

\usepackage{color}

\usepackage{graphicx}
\graphicspath{{figures/}}

\newcommand{\RR}{\right}
\newcommand{\LL}{\left}
\newcommand{\m}{\mathrm}
\newcommand{\dg}{\dagger}
\newcommand{\eref}[1]{Eq.~(\ref{#1})}
\newcommand{\fref}[1]{Fig.~\ref{#1}}
\newcommand{\puoli}{\frac{1}{2}}

\begin{document}

\title{Squeezing of quantum noise of motion in a micromechanical resonator}


\author{J.-M. Pirkkalainen$^{1}$}
\author{E. Damsk\"agg$^{1}$}
\author{M. Brandt$^{1}$}
\author{ F. Massel$^{2}$}
\author{M.~A.~Sillanp\"a\"a$^{1}$}
\thanks{Mika.Sillanpaa@aalto.fi}
\affiliation{$^1$Department of Applied Physics, Aalto University, PO Box 11100, FI-00076 Aalto, Finland. \\
$^2$University of Jyv\"askyl\"a, Department of Physics, Nanoscience Center, University of Jyv\"askyl\"a, PO Box 35 (YFL) FI-40014 University of Jyv\"askyl\"a, Jyv\"askyl\"a, Finland.}

\begin{abstract}



A pair of conjugate observables, such as the quadrature amplitudes of harmonic motion, have fundamental fluctuations which are bound by the Heisenberg uncertainty relation. However, in a squeezed quantum state, fluctuations of a quantity can be reduced below the standard quantum limit, at the cost of increased fluctuations of the conjugate variable. Here we prepare a nearly macroscopic moving body, realized as a micromechanical resonator, in a squeezed quantum state. We obtain squeezing of one quadrature amplitude $1.1 \pm 0.4$ dB below the standard quantum limit, thus achieving a long-standing goal of obtaining motional squeezing in a macroscopic object.
  \end{abstract}
  
\maketitle




The motion $x(t) = X_1(t) \cos (\omega_m t) + X_2(t) \sin (\omega_m t)$ of a harmonic oscillator having the natural oscillating frequency $\omega_m$ can be described by the quadrature amplitudes $X_1$ and $X_2$ which have slow fluctuations. 
The fluctuations, presented in units of the quantum zero-point fluctuation amplitude $x_{\m{zp}}$, satisfy the Heisenberg uncertainty relation $\Delta X_1 \Delta X_2 \geq 1$. One of the two can be prepared (=squeezed) below the value 1, at the expense of increased fluctuations in the other quadrature. In optics, squeezing of laser light was observed in early 80's \cite{Walls1983Squeeze,Squeeze1985}, not long after the possibility was realized.




It has been a formidable challenge to obtain squeezing in the motional state of a macroscopic object. The possibility of squeezing in the oscillations of massive gravitational antennae has been hypothesized a long time ago \cite{GravSqu1979,GravSqu1983}, but technological limitations are too severe for experimental realization. Other motional quantum-mechanical phenomena, on the other hand, have recently been experimentally demonstrated  \cite{ClelandMartinis,LehnertEnta2013} in micromechanical resonators. The latter systems are nearly macroscopic in physical size, and therefore they provide an ideal test system for treating the borderline between quantum and classical. Of particular interest for these studies has been the cavity optomechanics setting coupling electromagnetic cavity mode and the oscillator motion \cite{OptoReview2014}. Output of squeezed light \cite{Atom2012Sq,Painter2013Sq,Regal2013Sq} was recently observed, but this does not yet imply the oscillator mode is squeezed.


    
  \begin{figure*}[htp]
 \includegraphics[width=0.9\linewidth]{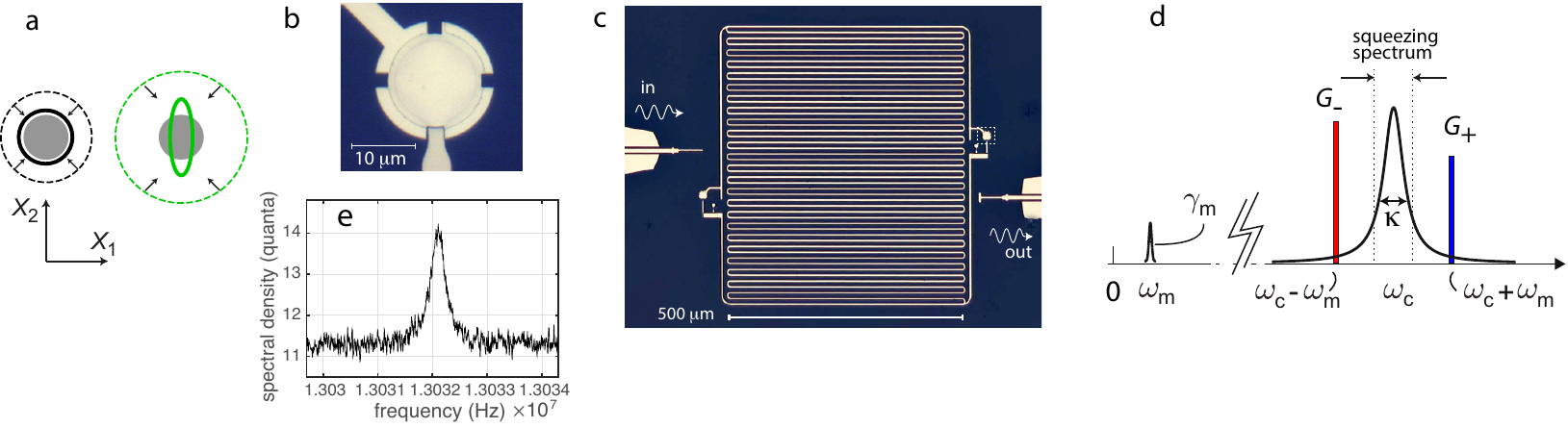}
  \caption{\emph{Setup of the microwave optomechanical squeezing experiment}. (a), Idea of dissipative squeezing. The initial fluctuation amplitudes of the quadratures are marked with dashed lines, and the final ones with solid lines. The green circles denote the quantum ground state. In sideband cooling (left), the initial fluctuations uniformly cool towards the ground state. Right, in a suitably engineered system, cooling can be quadrature-dependent, hence leading to one quadrature becoming squeezed. (b), (c), Optical micrographs of the micromechanical device and of the cavity. There are two drum resonators connected to cavity, however, only one of them (within the dashed rectangle) is operated. The cavity is asymmetrically coupled to the input port (left), and to the output port (right) for transmission measurements. (d), The frequencies involved in the scheme. The cavity is pumped by two nonequal-amplitude microwave tones at the sideband frequencies $\omega_\pm = \omega_c \pm\omega_m$. (e) Example of the thermal motion signal measured at the refrigerator base temperature, with the coupling $G_-/2\pi \simeq 1.3$ kHz.}
  \label{fig1}
\end{figure*}


Here we report the first realization of squeezing of the motional state of a nearly macroscopic body, realized as a micromechanical resonator measuring $15$ microns in diameter. We utilize the novel idea of dissipative squeezing \cite{ClerkSq2013,Meystre2013Sq,Home2015Sq}, see \fref{fig1} a, where the system is allowed to cool towards a squeezed low-energy state. This method has the great advantage of being able to create unconditional squeezing in the steady-state. This contrasts to many other plausible methods of squeezing generation  \cite{Blencowe2000Sq,Schwab2008,Marquardt2008Sq,Tian2008Sq,Zoller2009Sq,Bowen2011Squ,Polzik2015Sq}. Our approach is closely related to the quantum nondemolition measurements \cite{Schwab2010QND,Schwab2012instab,Schwab2014QND} which, however, are not able to generate true squeezing without feedback. At this point we mention that classical squeezing of thermal noise 
is routinely observed in mechanical systems \cite{Rugar1991,Buks2007Sq,YamaguchiSqu2014}.

The mechanical element (\fref{fig1} b) is a drum resonator basically similar to Refs.~\onlinecite{Teufel2011a,Teufel2011b}. The 120 nm thick Al drum fabricated on a quartz substrate is connected through a narrow 70 nm vacuum gap to one end of the cavity. The latter consists (\fref{fig1} c) of a meandering superconducting Al strip. This kind of microwave optomechanical systems are conveniently described using a model involving lumped electromagnetic elements. The interaction between the microwave cavity and the mechanical vibrations is given by the energy $g_0 n_P x$, where $n_P$ is the number of photons externally applied by microwave pump(s), and $g_0$ is the radiation-pressure coupling energy. Because usually $g_0$ is much smaller than other energy scales, a large $n_P \gg 1 $ is needed to effectively enhance the interaction up to a value $G = g_0 \sqrt{n_P}$. If the pump is applied at the frequency $\omega_- = \omega_c - \omega_m$ (the red sideband), the physics leads to sideband cooling of the mechanical vibrations, possibly down to the quantum ground state \cite{Teufel2011b,AspelmeyerCool11}. 

The scheme of Ref.~\cite{ClerkSq2013} requires two pump microwaves, one applied at the red sideband, and the other at the blue sideband frequency $\omega_+ = \omega_c + \omega_m$ (\fref{fig1} d). The two pertinent effective couplings are called $G_-$ and $G_+$, respectively. This setup can be described as sideband cooling of a Bogoliubov (BG) mode \cite{ClerkSq2013,supplement}, which in laboratory frame corresponds to cooling the mechanical mode towards a squeezed vacuum state. The BG mode is defined by the annihilation operator $\beta \equiv b \cosh r + b ^\dg  \sinh r$, obtained from the creation and annihilation operators $b^\dg$, $b$ of the mechanical resonator. An arbitrary squeezing ratio $r$ can be selected by tuning the ratio of the two pumps: $\tanh r = G_+ / G_- $. The BG mode - cavity system is described by the Hamiltonian $ H = G_\beta \LL( a^\dg  \beta + a \beta^\dg \RR)$ with the coupling energy $G_\beta = \sqrt{G_-^2 -G_+^2}$. Here, $a^\dg$, $a$ are the creation and annihilation operators of the cavity.  Although a value $G_- / G_+ \gtrsim 1$ would give rise to a high squeezing ratio, the effective coupling needed for cooling this mode would vanish. Hence, an optimum value is typically around $G_- / G_+ \sim 1.5 $.

The measurements are carried out inside a Bluefors dry dilution refrigerator in the temperature interval 7 mK ... 200 mK. The cavity is first characterized having the frequency $\omega_c/2\pi \simeq 6.9004$ GHz, and coupled to the transmission lines with the decay rates $\kappa_{{Ei}}/2\pi \simeq 50 \pm 5$ kHz, $\kappa_{{Eo}}/2\pi \simeq 270 \pm 30$ kHz through the input and output ports, respectively. The internal losses are characterized by the rate $\kappa_I/2\pi \simeq 330 \pm 40$ kHz, and the total cavity damping rate is $\kappa = \kappa_{{Ei}} + \kappa_{{Eo}} + \kappa_I  \simeq (2\pi ) \cdot 650 \pm 10$ kHz. We operate in the good cavity limit $\omega_m/\kappa \simeq 20  \gg 1 $, a prerequisite for efficient sideband cooling and squeezing generation. 

The mechanical resonator is first characterized using a single pump tone at the red sideband. We choose very low pump powers such that the cavity back-action damping rate $\gamma_- = 4 G_- ^2 /\kappa$ is much smaller than the  intrinsic linewidth $\gamma_m$ of the mechanics. The emission at the cavity frequency then shows the usual thermal motion peak at the motional sideband at a frequency $\omega_m/2\pi \simeq 13.032$ MHz above the pump. We obtain $\gamma_m/2\pi \simeq 330$ Hz corresponding to the $Q$-value $Q_m \simeq 3.9 \times 10^4$ from the data as in \fref{fig1} e ($\gamma_- \sim 12$ Hz was subtracted from the fit result).

    \begin{figure}[htp]
 \includegraphics[width=0.95\linewidth]{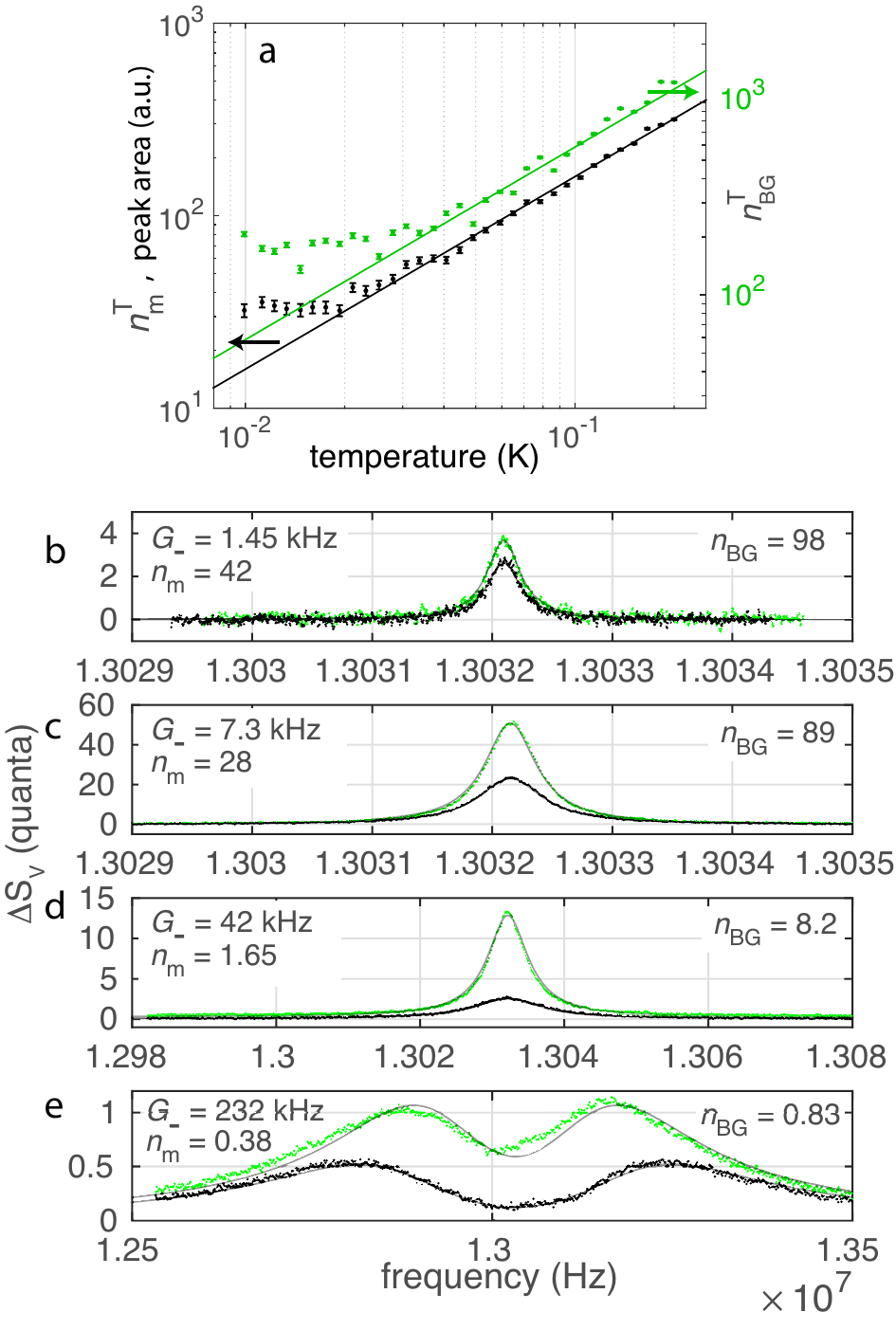}
  \caption{\emph{Cooling the Bogoliubov mode}. In all the panels, black color refers to regular  sideband cooling (i.e., $G_+ =0 $) used as calibration, whereas green refers to the BG mode experiment. (a), Thermalization in equilibrium. The left scale gives $n_m^T$, as well as the area of both the sideband cooling peaks, and of the BG mode peaks (both in a.u.), whereas the right hand side scale is $n_{BG}^T$. (b)-(e), Output spectrum during sideband cooling (black), or under BG mode conditions (green). The solid lines are theoretical curves. The $n_m$ values refer to the $G_+ =0 $ case.}
  \label{fig2}
\end{figure}

An important benchmarking for cavity optomechanical experiments is how well the mechanical mode thermalizes to the temperature $T$ of the refrigerator. Here we observe the linear temperature dependence expected (as seen in \fref{fig2} a) in equilibrium, $k_B T = n_m^T  \hbar \omega_m$ down to $\simeq$ 25 mK. Here, $n_m^T$ is the equilibrium thermal phonon number defined accordingly. In what follows, we operate at the minimum $T$ where we know that the mechanical mode is at 25 mK corresponding to $n_m^T \simeq 40$ at low pump powers.

We proceed with a series of further calibrations on the way towards demonstrating squeezing. Next we perform a regular sideband cooling experiment ($G_+ = 0$), see e.g.~Refs.~\cite{Aspelmeyer2006cool,Kippenberg2006cool,Schwab2010,multimode2012,GrapheneOpto}. For calibrating $G_-$ versus generator power, we study the peak width $\gamma_m + \gamma_- $ as a function of power at modest values of $G_- \ll \kappa$. The most critical step, which also will account for most of the final imprecision, is to calibrate the system gain at the detection side. Each sideband cooling spectrum (about 20 curves at different power) are simultaneously fitted to theory with the same gain, using the $G_-$ and $n_m^T$ just calibrated. For details, see the Supplementary \cite{supplement}. We show examples of the sideband cooling spectra by the black symbols in \fref{fig2} b-e, overlaid with theoretical predictions from the standard formalism using input-output theory \cite{supplement}. For the plot, we have subtracted a large background level due to the amplifier noise, hence displaying only the signal part due to the sample. We also find that the mechanical mode cools down to a thermal occupation $n_m \simeq 0.38$ ($n_m = 0$ corresponds here to the ground state). The double peak seen in \fref{fig2} e signifies the onset of the strong-coupling regime when $G_- \sim \kappa$. The data is plotted in dimensionless units (quanta) which is the natural unit from the theory point of view (W/Hz units are obtained by multiplying by $\hbar \omega_c$).





Given that we can cool the drum motion close to the ground state provides a promising starting point for creating squeezed motional states. We switch on the blue pump while keeping the red on, obeying $G_- > G_+ $ for stability. This creates a certain BG mode depending on the ratio $G_- / G_+$. In order to ascertain which BG mode we have, we calibrate the input line attenuations separately for both pump generators. We select $\gamma_\pm = 4 G_\pm ^2 /\kappa \ll \gamma_m$ and adjust the powers such that we obtain equal response due to either pump. The imprecision is estimated to be $\pm 0.2$ dB which is also the imprecision for constructing a given BG mode.



Next we discuss a specific BG mode obeying $G_- / G_+ \simeq 1.52$ which is expected to represent a choice close to optimum. At the lowest pump powers, the back-action cooling is negligible and we reveal the bare BG mode undressed from the cavity. Under this condition, the equilibrium BG mode occupancy $n_{BG}^T$ is expected to follow a linear temperature dependence same way as the mechanical mode, but with an elevated temperature \cite{supplement}. More relevant than $n_{BG}^T$, however, is the agreement of the spectra with theoretical prediction, which here is connected to the area under the Lorentzian BG mode peak. We test this in \fref{fig2} a and observe an excellent agreement to the theory. The green solid line is an expectation based on the calibrated $G_- / G_+$ ratio, and on the $G_+ = 0$ data.




Next, we increase the effective couplings. In order to mitigate possible gain drifts, we repeat a sequence of short measurements of the amplifier noise background, sideband cooling, and BG data. Plotted in conjunction with the sideband cooling data, in \fref{fig2} b-e we display the BG mode spectra in green. The theoretical plots show a good agreement to the data.  In \fref{fig2} e, the BG mode curve is slightly shifted to the left probably because of an energy-dependent shift in the cavity frequency. When using fixed pump frequencies as here, the cavity can become slightly detuned at certain pump powers.
Here we have used as adjustable parameters the bath temperatures of both the cavity and the mechanics. Both baths heat up with the pump powers, for the BG mode up to $n_{I} \simeq 0.74$, $n_{m}^T \simeq 86$ (instead of $n_{BG}^T$, we prefer to quantify the bath with $n_{m}^T$) in \fref{fig2} e, which is attributed to dielectric heating. These factors together set the limits for the cooling as well as squeezing. As one can anticipate, the baths heat up more if we apply both pumps instead of only one. 

In \fref{fig2} b-e we label the fitted values of the BG mode occupation $n_{BG}$ which is a quantity analogous to $n_m$ in case of regular sideband cooling. Although $n_{BG}$ does not directly correspond to a physical temperature, a value $n_{BG} < 1$ as obtained in \fref{fig2} e is indicative of squeezing. We can more precisely infer the amount of squeezing by evaluating the quadrature variances \cite{supplement}. This way, we obtain that the mechanical mode is squeezed below the vacuum level, in \fref{fig2} e by $\simeq 1.1$ dB.


%
      \begin{figure*}[htp]
 \includegraphics[width=0.8\linewidth]{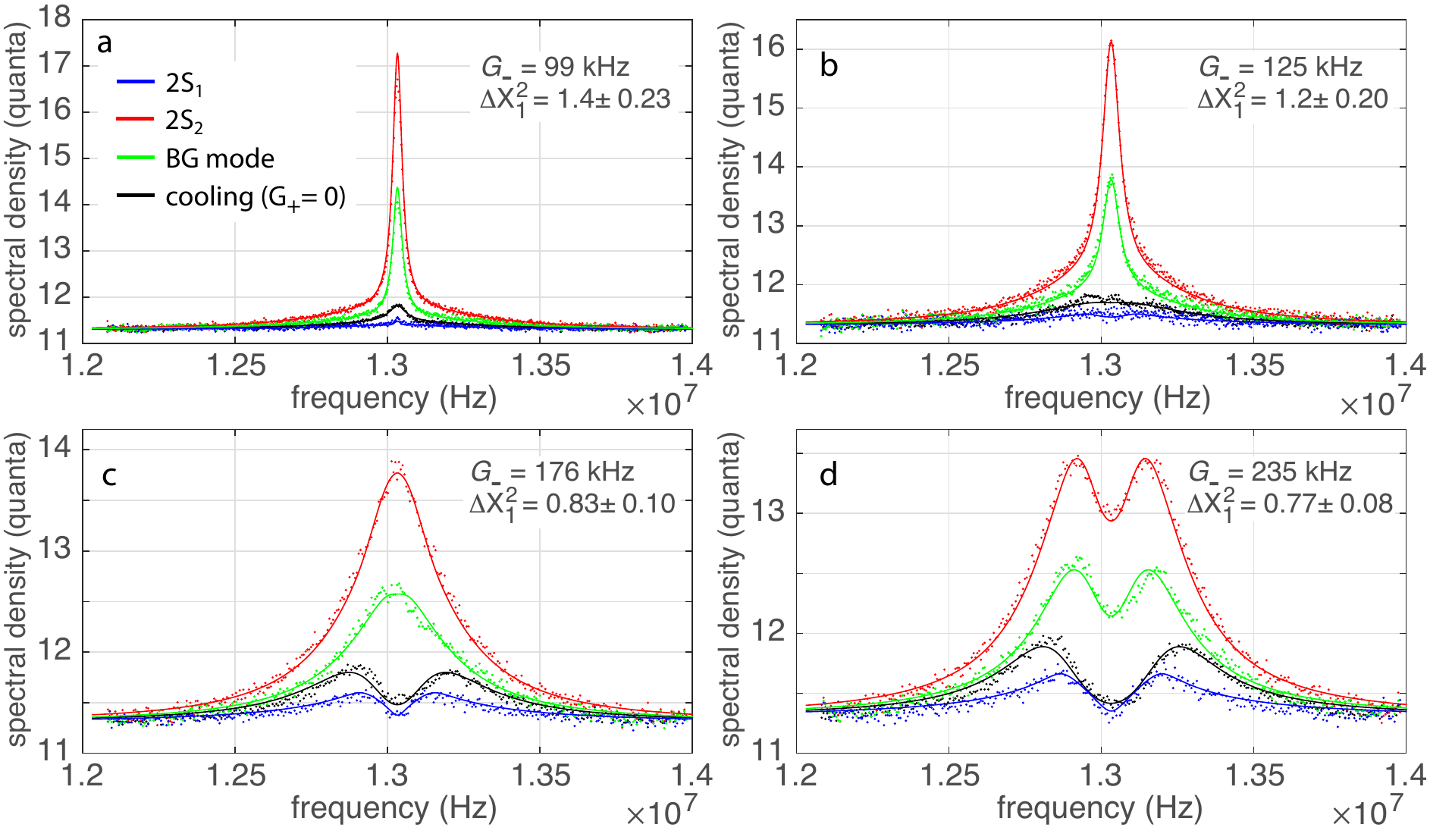}
  \caption{\emph{Squeezing inferred from the quadrature spectra}. In all panels, blue and red refer to the cold and hot quadratures $X_1$ and $X_2$, respectively. Black and green refer to the regular sideband cooling, and the BG mode, respectively. The solid lines are theory curves. The pump powers are increased from (a) to (d) as marked in the panels while $G_-/G_+ \simeq 1.43 $ ratio is kept constant. The variances $\Delta X_1^2$ are marked, and a value less than one implies squeezing below vacuum. The amplitudes of parametric modulation to the cavity are $\epsilon_c/2\pi \simeq 35, 48, 49, 56$ kHz  from (a) to (d).}
  \label{fig3}
\end{figure*}

The best verification of squeezing comes from the quadrature spectra which amount to tomography of the state. We digitally mix down the signal using the center frequency $\omega_c$ of the pumps as a local oscillator (LO), hence making homodyne detection. The quadrature spectra show strong dependence on the LO phase. We identify the minimum emission as the $X_1$ quadrature, and the maximum, offset by 90$^{\m o}$, as $X_2$. In \fref{fig3} we show the corresponding quadrature spectra $S_1$ and $S_2$  together with the total spectrum (the BG mode) and the cooling spectrum. We plot the quantities $2 S_1$ and $2 S_2$  for more conveniently presenting them together with the other two curves. The theoretical predictions overlaid on the data show an excellent agreement. In the best case of \fref{fig3} d, the motion of the mechanical resonator has been squeezed about 1.1 dB below the Heisenberg limit. When varying the LO phase, we also observe an excellent agreement to theory (\fref{fig4}). The bath temperatures are found to be slightly enhanced over the previous dataset in \fref{fig2} \cite{supplement}. We also find that this data agrees with a slightly shifted BG ratio $G_-/G_+ \simeq 1.43$ and which is attributed to drift, not directly measured, in the tunable filters at room temperature over the about one week after the calibration.



    \begin{figure}[htp]
 \includegraphics[width=0.8\linewidth]{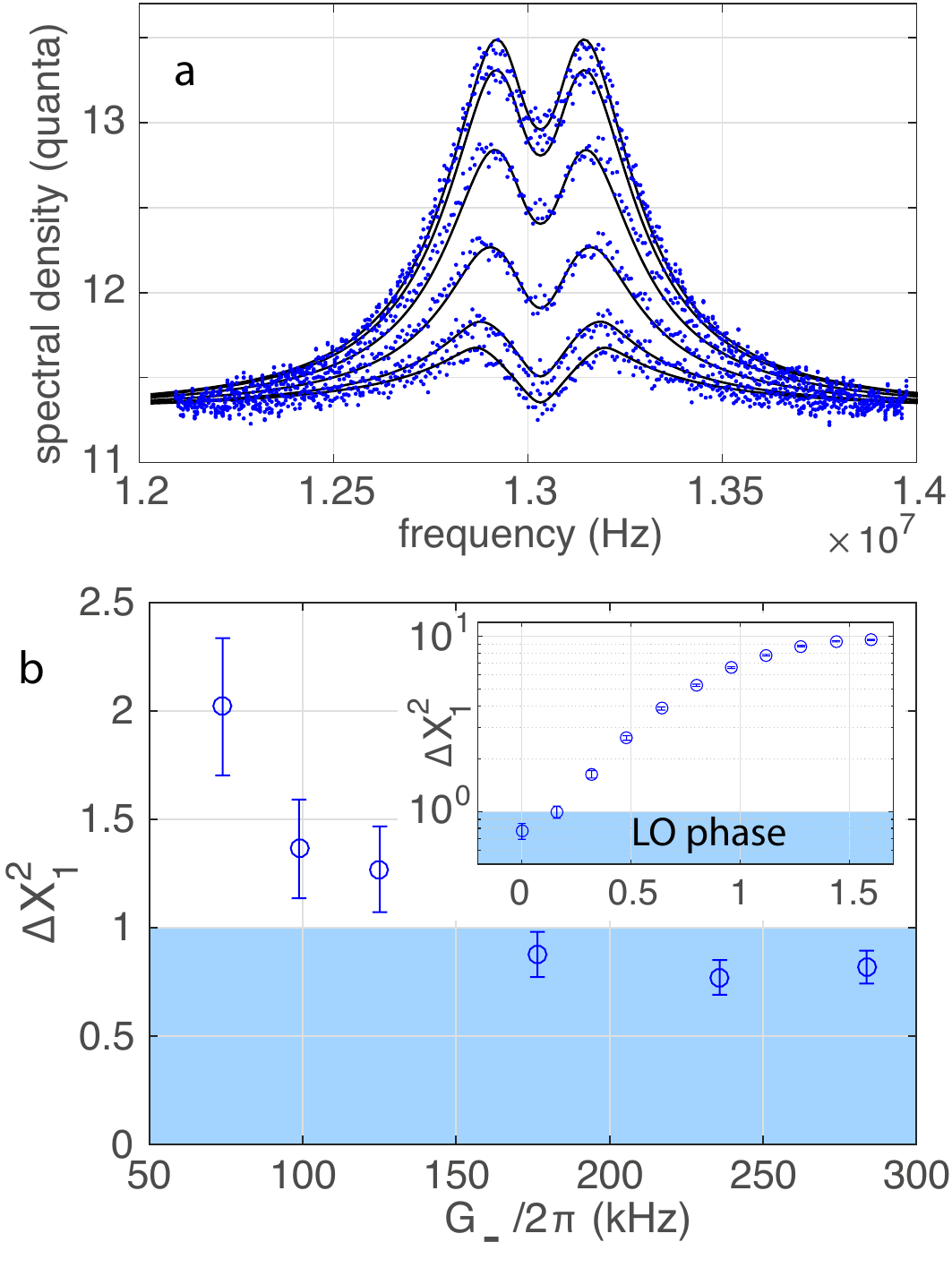}
  \caption{\emph{Tomography and final results}. (a), The quadrature data similar to Fig.~3 d, plotted at different LO phase values at $\pi/10$ steps from 0 to $\pi/2$, from bottom to top. (b), The $X_1$ quadrature variance as a function of pump power. The blue region signifies squeezing below the quantum limit. The inset shows the quadrature variance from the data in (a).}
  \label{fig4}
\end{figure}



We now discuss an issue which is critical for understanding the quadrature spectra, namely, parametric effects beyond the ideal optomechanical model. One can satisfactorily model the total BG mode spectrum (and get equal squeezing) without imposing any parametric modulation \cite{supplement}, however, the quadratures show much more phase dependence than predicted. In the scheme, mixing products of the pumps can appear both at $2\omega_m$ as well as $2\omega_c$, the frequencies most susceptible to cause parametric modulation via e.g.~thermal effects or nonlinearities \cite{Schwab2012instab,supplement}. Modulation of $\omega_c$ out of phase with respect to the pumps gives an excellent match to the quadratures with the values quoted in \fref{fig3}. In order to verify the existence of said parametric effect, we carried out a measurement where we substantially detuned both pumps from the sideband resonance, by $\sim \pm 6 \kappa$, such that the center frequency stays at $\omega_c$ \cite{supplement}. This high detuning essentially eliminates optomechanical phenomena, but a possible field at $2\omega_c$ remains. This way we measure a spectrum consistent with a parametrically modulated oscillator with the correct phase. A possible explanation is nonlinear dissipation in the cavity \cite{Bachtold2011NLdamp}, or a thermal effect. Although the parametric effects have a dramatic influence on the quadratures of the output spectrum, they only weakly affect the squeezing of the mechanics, in the present case we find a reduction by about 10 \%.



For the error analysis, we use a worst-case scenario of systematic errors from the calibrated parameters, and of direct statistical errors of the adjustable parameters. We find that the gain uncertainty is dominating. Because the shapes of the spectra are sensitive to most parameters, but squeezing is somewhat insensitive to any parameters in the present parameter region, we obtain a relatively small imprecision of $\sim 10$ \%. The final numbers are quoted in \fref{fig4} b.



In conclusion, we have achieved the reduction of the quantum fluctuations in a nearly macroscopic moving object below the "sound of silence" level by 1.1 dB in one quadrature. Following the discoveries of ground-state cooling \cite{ClelandMartinis,Teufel2011b,AspelmeyerCool11}, observation of zero-point fluctuations in mechanical systems \cite{Painter2012ZP}, and entanglement \cite{LehnertEnta2013}, our work goes forward to establish the reality of another fundamental quantum property in moving objects.

Note added: while finalizing the paper, we became aware of a similar experiment \cite{ScwabSqueeze}.




\bibliography{/Users/masillan/Documents/latex/MIKABIB}

\textbf{Acknowledgements} We would like to thank Pasi L\"ahteenm\"aki, Tero Heikkil\"a, Souvik Agasti, and Andr\'e Xuereb for useful discussions. This work was supported by the Academy of Finland (contract 250280, CoE LTQ, 275245) and by the European Research Council (240387-NEMSQED, 240362-Heattronics, 615755-CAVITYQPD). The work benefited from the facilities at the Micronova Nanofabrication Center and at the Low Temperature Laboratory infrastructure.


\newpage
\vspace{3.5cm}

\begin{widetext}
\vspace{1.5cm}

\noindent \Large{Supplementary Information for ''Squeezing of quantum noise of motion in a micromechanical resonator''}

\normalsize

\vspace{0.5cm}

\noindent J.-M. Pirkkalainen$^{1}$
E. Damsk\"agg$^{1}$
M. Brandt$^{1}$
 F. Massel$^{2}$
M.~A.~Sillanp\"a\"a$^{1}$\\

\noindent $^1$Department of Applied Physics, Aalto University, PO Box 11100, FI-00076 Aalto, Finland. \\
$^2$University of Jyv\"askyl\"a, Department of Physics, Nanoscience Center, University of Jyv\"askyl\"a, PO Box 35 (YFL) FI-40014 University of Jyv\"askyl\"a, Jyv\"askyl\"a, Finland.

\maketitle

\section{Experimental setup}

We first mention how the microwave optomechanical device can be simply described using lumped electromagnetic elements (see   \fref{uwOptoSchema2}). The motion of the drum changes its capacitance $C_{g}(x)$. The total capacitance of the cavity can be summed up as a constant $C$ and an $x$-dependent part $C_g(x)$. The frequency of this cavity hence is
\begin{equation}
\label{eq:fcuw}
\omega_c = \frac{1}{\sqrt{L (C + C_g(x))} } \,.
\end{equation}
Linearization gives the coupling 
\begin{equation}\label{eq:g1}
   g_0 =  \frac{\partial \omega_c}{\partial x} x_{\m{zp}} = \frac{\omega_c}{2C } \LL( \frac{\partial C_g}{\partial x } \RR) x_{\m{zp}}\; .
\end{equation}

Modeling the device layout using electromagnetic simulation software gives the equivalent parameters $C \sim 16$ fF, $L \sim 14$ nH. It is beneficial to get as low $C$ as possible in order to maximize coupling. Here the role of quartz substrate is critical because it has a low dielectric constant $\epsilon_r \sim 4$ as compared to sapphire or silicon ($\epsilon_r \sim 10 ... 12$). This difference amounts to a factor of two in $C$. 

A narrow vacuum gap $\sim 70 \pm 10$ nm between the drum and a bottom electrode deduced from the device structure corresponds to $\partial C_g/\partial x \sim 120$ nF/m, $x_{\m{zp}} \sim 5$ fm, and hence allows for a relatively large single-photon coupling energy $g_0 /2\pi \simeq 90 \pm 20$ Hz. We note that in the analysis, we are not directly using  $g_0$, but the calibrations and analysis is using the derived quantities $G_\pm$.

\begin{figure*}[htp]
 \includegraphics[width=0.9\linewidth]{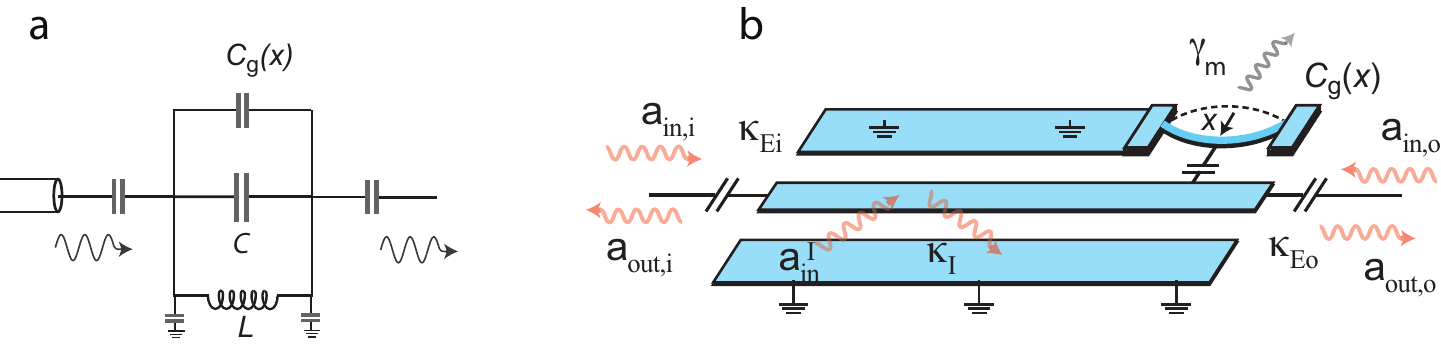}
  \caption{\emph{Schematics of the microwave optomechanical device}. (a), Model in terms of lumped electromagnetic elements. (b), Modeling scheme in the input-output theory. The cavity is exposed to noise from both the input (subscripts $i$) and output (subscripts $o$) couplers, as well as from the internal losses ($I$).}
  \label{uwOptoSchema2}
\end{figure*}


We use a pulse-tube powered Bluefors ''dry'' dilution refrigerator to cool the device down to about 7 mK.  Inside the cryostat, the incoming pump signal is attenuated at all temperature stages by a total of about 40 dB (including cabling) in order to dampen the Johnson noise from higher temperatures. We used relatively small-valued attenuators at the two lowest temperature stages in order to avoid heating the refrigerator by the pump microwaves. However, one can check that the input noise $n_i$, see \eref{incorrelators}, (considering an ideal setting) is only a vanishing amount above the vacuum level at $\sim$ 7 GHz. The entire cabling is represented in Fig.~\ref{cryosetup}. The signal from the sample is fed via mostly superconducting coaxial cables and two isolators into the low-noise amplifier at 4 Kelvin stage of the cryostat. A band pass microwave filter (BPF) is used to cut noise outside the isolator band. 

We will turn the discussion to the room-temperature instrumentation as shown in \fref{meassetup_squeezing2}. At the input side, the two pump tones from two low-phase noise microwave sources are combined with a power splitter. A tunable notch filter from Wainwright Instruments at the signal input at room temperature provides about 50 dB rejection ratio at the pump frequencies as compared to the center frequency $\omega_c$, and is meant to clean the phase noise of the pump sources. The phase noise would appear as an increased thermal noise $n_i$ to the cavity,  limiting the performance. Without the filter, indeed, there is noticeable heating of the cavity at the highest used pump powers.

\begin{figure}[htp]
 \includegraphics[width=9cm]{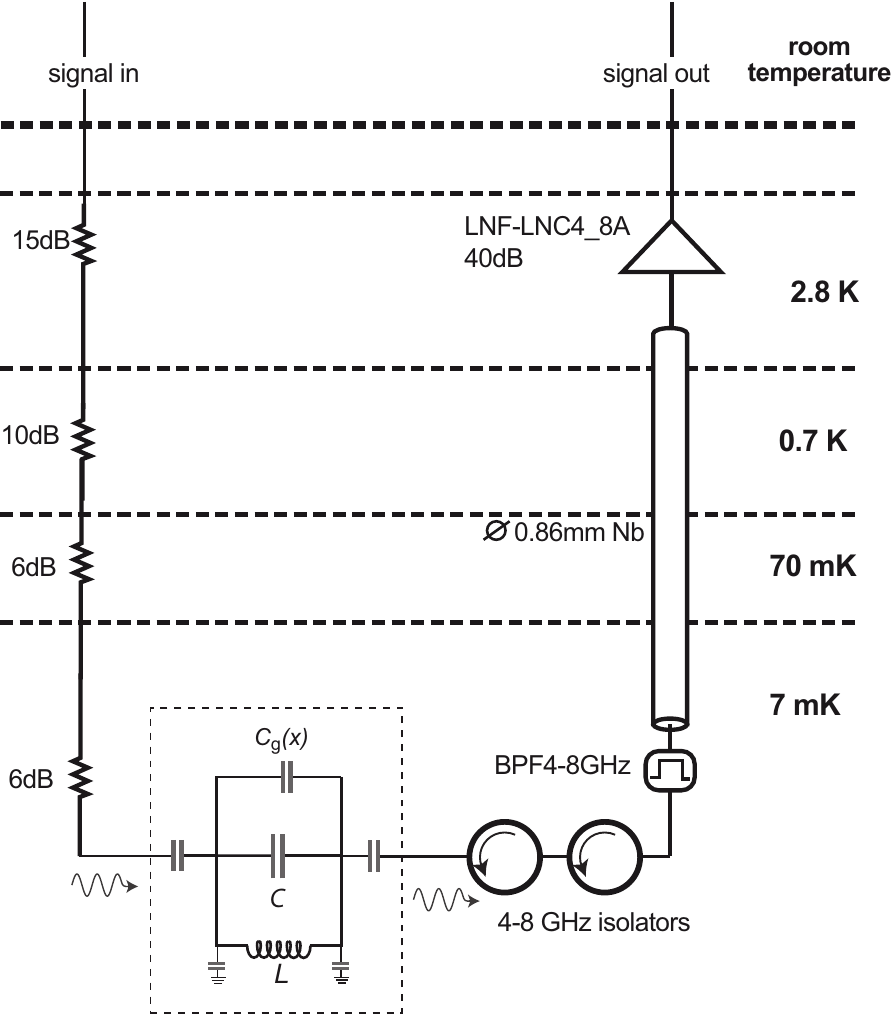}
  \caption{\emph{Setup for the cryogenic microwave measurements}.}
  \label{cryosetup}
\end{figure}


The input signal is further split, and one branch is used to create the local oscillator (LO) at the center frequency which is phase-coherent with both the two pumps. The branch for LO creation is marked in \fref{meassetup_squeezing2} as the shaded box. The signal is squared which creates a tone at $2 \omega_c$, then filtered around this frequency in order to block unintended tones at $2 \omega_c \pm 2 \omega_m$. The frequency prescaler ADF5000BCPZ then creates the desired tone at $ \omega_c$. It is attenuated to a suitable amplitude before being summed up to the signal from the sample, to be fed to the signal analyzer.

At room temperature, the signal coming from the sample is first further amplified by 30 dB, then band-pass filtered within the bandwidth of 4 MHz around the cavity frequency. This filtering both acts as a antialiasing filter for the digitizer, as well as attenuates the pump tones by 15-20 dB to avoid saturation of the next stages.

The vector network analyzer (VNA) is used to monitor the cavity response and hence enable for example correcting for occasionally observed drift in the cavity frequency.

We use Anritsu MS2830A signal analyzer which digitizes the datastream at 2 MHz bandwidth for the duration of approximately 1 second, and automatically makes the conversion into IQ data which is sent to computer. At the computer, we first extract the phase of the LO marker signal from the time-domain data at about 10 ms time resolution, and use this information, first of all, to make small corrections to the data due to phase drift. We construct the quadrature spectra as follows. The IQ data is converted into real-valued time-domain data as
$$
I(t) = |IQ(t)| \cos (\arg (IQ(t)))\,,
$$
$$
Q(t) = |IQ(t)| \sin (\arg (IQ(t)))\,.
$$
These quadratures are Fourier-transformed. We then construct a set of phase-rotated frequency-domain datasets as
$$
S_{V,\Theta}(\omega) \equiv I_\Theta(\omega) = \cos(\Theta) I(\omega) - \sin(\Theta) Q(\omega)\,.
$$
The final result of the processing is the quadrature spectrum $|S_{V,\Theta}(\omega) |^2$. 

The homodyne detection together with the fact that the peaks are centered at the LO frequency, cause that only one side of the \emph{quadrature} spectra peaks are observable. For convenience, we however plot the data mirrored in about the center frequency. In the plots the data points do not appear symmetrically situated, however, which is due to a moving average filtering applied. We cannot know if the original quadrature spectra are symmetric, however, the fact that the total output spectrum (the BG mode) is symmetric about the center frequency, as well as the theoretical prediction, indicate they are symmetric.

For obtaining the total spectra for the BG mode, or in case of sideband cooling, we directly process the spectrum as
$$
S_V(\omega) = \LL| \m{fft}[IQ(t)] \RR|^2 \,.
$$

All the instruments are locked into a common 10 MHz Rubidium frequency standard. We noticed that the Ru clock improves the phase stability by about 20 dB as compared to using a quartz clock, substantially easing the digital processing.


\begin{figure}[htp]
 \includegraphics[width=17cm]{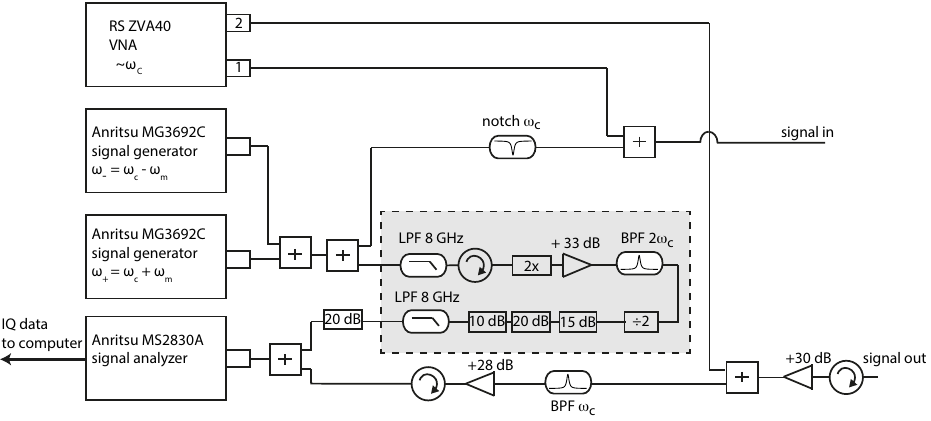}
  \caption{\emph{Setup for the room-temperature microwave measurements}. For discussion, see text.}
  \label{meassetup_squeezing2}
\end{figure}

It is important to prevent the LO signal at $\omega_c$ from entering the sample, because it would add a drive to the mechanics. We noticed that several loose connectors can be enough to allow for this tone to pass to the sample as a crosstalk through air. In this case we could see an enhanced temperature of the mechanics if the LO generation was switched on. In the final setup, this problem disappeared.

Since we are not using a nearly quantum-limited Josephson parametric amplifier, the signal-to-noise is often much smaller than one. This leads to the issue that data recording takes long time, which causes the signal level to be sensitive to drifts of the total gain of the amplification chain. Because the vertical scale of the output spectra needs to be known accurately, it is imperative to make sure gain drifts are taken care of. To this end, while taking the quadrature spectra, instead of averaging one curve for a long time, we go through measuring a set of curves each with a different pump power. At a time, we record one curve for about 2 seconds. Before switching to the next one, we run two calibrations. The first is a measurement of the background, namely, switching all pumps off and recording the amplifier noise spectrum over the same band. This allows neutralizing gain drifts. The second is the regular sideband cooling which allows keeping track of possible drift in the input attenuation and thermalization. 


\section{Theoretical modeling}

The cavity is described by the annihilation and creation operators $a$ and $a^\dg$, and the mechanical resonator similarly by $b$ and $b^\dg$. We use the quadratures of the mechanics
\begin{equation}\label{X1X2}
\begin{split}
& X_1 = b^{\dg} +b \\
& X_2 = i \LL(b^{\dg} -b \RR) \,,
     \end{split}
\end{equation}
or the other way round:
\begin{equation}
\begin{split}
          &b=  \puoli \LL(X_1 + iX_2\RR) \\
     &b^{\dg} = \puoli \LL(X_1 - iX_2\RR) \,.\\
     \end{split}
\end{equation}
Involving an arbitrary quadrature angle $\Theta$, one can write
\begin{equation}\label{Xtheta}
\begin{split}
& X_{\Theta} = b^{\dg} \exp (i \Theta )+b \exp (-i \Theta) \,.
     \end{split}
\end{equation}
%
%
%
Similar relations of course hold for the cavity operators which we do not write explicitly.

\subsection{Basic Hamiltonian}
\label{sec:ideal}

The Hamiltonian of the system consisting of the radiation-pressure coupled cavity and the mechanical resonator is
\begin{equation}
\label{eq:hsqueeze}
H =  \omega_c A^{\dg} A+  \omega_{m}b^{\dg}b+  g_0 A^{\dg} A x \,.
 \end{equation}
The uppercase cavity operators $A$ and $A^{\dg} $ involve both the quantum fluctuations described by the operators $a$ and $a^{\dg} $, as well as the rotating classical fields due to the two pumps:
\begin{equation}
\label{eq:clerkpumps}
A = \alpha_- \exp (-i \omega_- t) + \alpha_+ \exp (-i \omega_+ t) + a \,.
\end{equation}
The quantities $\alpha_+$, $\alpha_-$ are the classical intra-cavity amplitudes. We can, without loss of generality, suppose that they are both real. A phase difference between the amplitudes would only shift the phase reference of the quadratures. We define the detunings of both pump frequencies from the cavity:
\begin{equation}
\label{eq:deltapm}
\Delta_\pm = \omega_\pm - \omega_c
\end{equation}
In the experiment we operate very close to the exact sideband co-resonance condition $\Delta_\pm = \pm \omega_m $, however, for the modelling, we relax this assumption and allow basically arbitrary $\Delta_\pm$.

We transform to a rotating frame with
\begin{equation}
U = \exp (i \omega_c'  t  a^{\dg}  a)\exp (i \omega_m' t  b^{\dg}  b)  \,.
 \end{equation}
At the sideband co-resonance, the arbitrary frequencies $\omega_c'$ and $\omega_m'$ equal the cavity and mechanical frequencies, respectively.

In accord to usual practice, we neglect higher-order terms in the fluctuation operators in what follows. This amounts to linearization owing to $|\alpha_\pm| \gg a$. The coupling term becomes
\begin{equation}
\label{eq:couple1}
\begin{split}
& G_-^* b \exp (i (-\omega_c + \omega_- - \omega_m') t) a + G_+^*b \exp (i (-\omega_c' + \omega_+ - \omega_m') t) a + \\
&+ G_- b \exp (i (-\omega_- + \omega_c' - \omega_m') t) a^\dg + G_+ b \exp (i (-\omega_+ + \omega_c' - \omega_m') t) a^\dg + \\
& +G_-^* b^\dg \exp (i (-\omega_c' + \omega_- + \omega_m') t) a + G_+^*b^\dg \exp (i (-\omega_c' + \omega_+ + \omega_m') t) a + \\
&+ G_- b^\dg \exp (i (-\omega_- + \omega_c' + \omega_m') t) a^\dg + G_+ b^\dg \exp (i (-\omega_+ + \omega_c' + \omega_m') t) a^\dg  \,.
\end{split}
 \end{equation}
Here, the effective couplings are
\begin{equation}
\begin{split}
& G_\pm \equiv g_0 \alpha_\pm \\
\end{split}
 \end{equation}
We select $ \omega_c' = ( \omega_+ + \omega_-)/2$, and $ \omega_m' =  ( \omega_+ - \omega_-)/2$, simplifying \eref{eq:couple1} to
\begin{equation}
\begin{split}
& G_-^* b^\dg  a  + G_+^*b  a + G_- b a^\dg + G_+ b^\dg a^\dg + G_+ b a^\dg \exp (-2i\omega_m' t)  + \\
& + G_-^* b a \exp (-2i \omega_m' t) + G_+^*b^\dg  a \exp (2i \omega_m' t) + G_- b^\dg a^\dg \exp (2 i \omega_m' t)  \,.
\end{split}
 \end{equation}
We are working in the good-cavity limit $\omega_m/\kappa \simeq 20  \gg 1 $, and hence we neglect the terms rotating at $2 \omega_m'$. Relatively close to the sideband co-resonance, this becomes $\sim 2 \omega_m$, and the rotating terms are suppressed by nearly 40 dB.
We obtain that the Hamiltonian of the double-pumped system can be written as a beam-splitter between the cavity and the Bogoliubov modes:
\begin{equation}
\label{eq:bghamilt}
\begin{split}
H &=  -\LL( \frac{ \Delta_+  + \Delta_-}{2} \RR) a^{\dg} a+ \LL( \omega_m  - \frac{ \Delta_+  - \Delta_- }{2} \RR) b^{\dg} b  + a^\dg \LL( G_- b +G_+ b^\dg \RR) + a \LL( G_-^* b^\dg +G_+^* b \RR) = \\
&= \Delta_S a^{\dg} a+ \Delta_A b^{\dg} b + G_\beta \LL( a^\dg  \beta + a \beta^\dg \RR) \,,
 \end{split}
 \end{equation}
where the symmetric $\Delta_S$ and asymmetric $\Delta_A$  detunings were defined, and
\begin{equation}
\begin{split}
&G_\beta = \sqrt{G_-^2 -G_+^2}\\
&\beta =  b \cosh r + b ^\dg  \sinh r \\
& \tanh r = G_+ / G_- \,.
 \end{split}
 \end{equation}

The temperature of the BG mode is given as
\begin{equation}
\begin{split}
&n_{BG} \equiv \beta^\dg \beta = b^\dg b u^2 + (b^\dg)^2 uv + b^2 uv + b b^\dg v^2
= b^\dg b u^2 + (b^\dg)^2 uv + b^2 uv + (b^\dg b +1) (u^2 - 1) = \\
&= 2 b^\dg b \cosh^2 r -b^\dg b+ \cosh r  \sinh r \LL( (b^\dg)^2  + b^2 \RR)  +\sinh^2 r \,.
     \end{split}
\end{equation}
At low effective couplings, the phase-sensitive correlations vanish, and we get
\begin{equation}
\begin{split}
&n_{BG}  =b^\dg b  \LL( \cosh^2 r +\sinh^2 r \RR)  +\sinh^2 r  \,.
     \end{split}
\end{equation}

\subsection{Parametric modulation}

On top of the ideal model in section \ref{sec:ideal}, we consider two additional effects, namely, a possible parametric modulation to the mechanics or to the cavity. We show below that the latter is critical for explaining the \emph{quadrature} spectra, whereas the former does not play a significant role. The former was investigated in detail by Suh \emph{et al.~}[23], who found that it can be a serious limitation in the double pump scheme if it is strong enough to cause parametric instability.

In general, the double pump scheme is prone to issues due to unintended parametric modulation. Said briefly, this is because mixing products of the pumps at $\omega_c \pm \omega_m $ can appear both at $2\omega_m$ as well as $2\omega_c$, the frequencies most susceptible to cause parametric effects. 

\subsubsection{Mechanics}

In the scheme, there is an unavoidable parametric modulation resulting from the second-order optomechanical coupling with the coupling coefficient $g_2 = \frac{\partial^2 \omega_c}{\partial x^2} = \frac{\omega_c}{2 C}\frac{\partial^2 C_g}{\partial x^2} $.

The Hamiltonian \eref{eq:hsqueeze} is added with
\begin{equation}
\label{eq:g2}
\begin{split}
 &H_2 = \puoli  g_2 A^{\dg} A x^2= \puoli  g_2 A^{\dg} A \LL(b^2 + 2 b^\dg b + (b^\dg)^2 \RR) \,.
        \end{split}
 \end{equation}
By the ansatz in \eref{eq:clerkpumps} in the rotating frame, this becomes in the RWA
%
%
%
\begin{equation}
\begin{split}
 &\hat{H}_2 =  g_2  \LL(  |\alpha_-|^2 +|\alpha_+|^2 \RR) b^\dg b + \puoli  g_2 \alpha_- \alpha_+b^2 + \puoli  g_2 \alpha_+ \alpha_- (b^\dg)^2  = G_2 b^\dg b + \puoli \varepsilon_m b^2 + \puoli  (b^\dg)^2 \,,
        \end{split}
 \end{equation}
where
\begin{equation}
\begin{split}
     & G_2^\pm \equiv \sqrt{ g_2 } \alpha_\pm = \sqrt{ g_2 } G_\pm /g_0  \\
     & G_2 \equiv (G_2^+)^2 + (G_2^-)^2  \\  
     &\varepsilon_m = G_2^+ G_2^- = g_2 G_- G_+/g_0^2
\end{split}
\end{equation}
The strength of the parametric modulation is hence given by the quantity $\varepsilon_m$. In our scheme, $g_2/(2 \pi)^2 \sim 2 \times 10^{-5}$ Hz$^2$. We obtain that at the highest pump powers used in the experiment, $\varepsilon_m/2\pi ~ \simeq 300$ Hz which is similar to the mechanical damping rate, and hence looks like we are close to a parametric instability. However, we find that the sideband cooling interaction counteracts parametric modulation, and that the mechanical parametric modulation has a vanishing effect for either the quadrature temperatures or output spectra for $\varepsilon_m/2\pi \lesssim 10$ kHz.

Apart from the second-order coupling in \eref{eq:g2}, parametric modulation of the mechanical frequency could arise from thermal effects [23], with a substantially larger amplitude. We cannot qualitatively analyze this because we do not know the thermal picture well, however, we find numerically that the output quadrature spectra would be affected in a way which is inconsistent with the measurement (see section \ref{sec:diffparam}). The peaks in the spectra would sharpen up from the ideal case and would sometimes show a multi-peak structure.  Hence, parametric modulation of the mechanical frequency in the experiment is rather well excluded.

\subsubsection{Cavity}


In the ideal model together with  \eref{eq:g2}, parametric modulation of the cavity, or fields rotating at $2\omega_c$, do not arise. However, nonlinearities in the cavity, such as a Kerr effect, $\omega_c \Rightarrow \omega_c \LL(1 + \frac{1}{4} K A^\dg A \RR) $,  would induce fields at $2\omega_c$. In the rotating frame we obtain
\begin{equation}
\begin{split}
     & H_{c2} = \frac{1}{4} K  (A^{\dg})^2 A^2 
     = \puoli K \alpha_- \alpha_+ (a^\dg)^2 + \puoli K\alpha^*_- \alpha^*_+ a^2 +K \LL(|\alpha_+| ^2 +| \alpha_-| ^2 \RR)  a^\dg a  \equiv \\
     &  \equiv  \puoli \varepsilon_c^* a^2+   \puoli \varepsilon_c (a^{\dg})^2  + K' a^\dg a 
\end{split}
\end{equation}
Here we also defined the complex modulation amplitude $\varepsilon_c$. Because the field amplitudes (\eref{eq:clerkpumps}) can be taken as real, $\varepsilon_c$ is real as well in case of the Kerr effect. However, as shown in section \ref{sec:diffparam}, the calibration and squeezing spectra indicate a purely imaginary $\varepsilon_c$ (phase shift $-\pi/2$) (section \ref{sec:paramcalib}). This type of phase shift is indicative of modulation of dissipation.

A possible explanation is nonlinear damping. The extra nonlinear damping term modeled as nonlinear coupling to a bath of oscillators [36] in the equation of motion for a $\ddot{X_1}$  equals $- \eta X_1(t)^2 \dot{X_1}$ with a real-valued proportionality constant $\eta$. For the equation of motion for the annihilation operator in the lab frame, this corresponds to the term $ \puoli \eta (A^2 + 2 A^\dg A + (A^\dg)^2) (A^\dag -  A)$ added to the rhs. In the present pump scheme, this term becomes $-\eta \alpha_- \alpha_+ a^\dg$ which corresponds to the required a purely imaginary $\varepsilon_c$ in the equations of motion, \eref{eqmotsq}. As typical of microfabricated cavities, we indeed observe power-dependent dissipation, but we have not been able to sort out the contributions due to different mechanisms, such as two-level system damping and breaking of superconductivity at high currents. This would require the understanding, for example, of the time scales of the different processes.


%
%

\subsection{Input-output modeling}

\label{sec:supplio}

The model for the transmission setup is depicted in   \fref{uwOptoSchema2}. The following modeling is basically standard input-output theory. The asymmetric cavity has the input side where the pumps are applied, with the corresponding input coupling rate $\kappa_{Ei}$. Energy inside the cavity can dissipate in three separate channels. One is through the input coupler. Another channel is the cavity internal losses at the rate $\kappa_{I}$, and the third is the output coupler with the rate $\kappa_{Eo} \gg \kappa_{Ei}$. The latter property enables us to catch nearly twice the signal as compared to a symmetric cavity, important for the integration time needed in the measurements.

The input fields to the cavity are the standard input-output theory quantum fields described by the operators $a_{in,i}$, $a_{in,o}$, $a_{in,I}$ for the input and output couplers, and internal losses, respectively. They have the frequency-domain correlators ($n = i,o,I$)
\begin{equation}
\label{incorrelators}
\begin{split}
     &  \langle a_{in,n}^{\dg}(\omega) a_{in,n}(-\omega) \rangle = n_{n} (\omega_c)  \\
     &  \langle a_{in,n}(\omega) a_{in,n}^{\dg}(-\omega) \rangle = (1+n_{n}(\omega_c))  \,,\\
     \end{split}
\end{equation}
other correlators being zero.

The three baths in general have unequal temperatures $T_n$ which give the individual Bose factors
\begin{equation}
\label{bose}
n_{n}(\omega_c) = \LL( \exp(\hbar \omega_c /k_B T_n )-1 \RR)^{-1}\,.
\end{equation}

The internal losses of the mechanics are due to the field $b_{in}$ which has similar properties as in \eref{incorrelators}, \eref{bose}, with the replacements $\omega_c \Rightarrow \omega_m$, $n_n \Rightarrow n_m^T$.

We suppose the $i$ and $o$ baths are at zero temperature which is standard assumption. We also have carefully taken care to isolate excess noise from both these ports. The fact that the thermal calibration and sideband cooling calibration work well without a dip arising from such noise supports this assumption. The cavity internal bath, however, rises up to $n_I \sim 0.9$ at the highest pump powers.


The equations of motion, including for completeness the parametric modulations to both the mechanics and cavity, as well as pump detunings, are
\begin{equation}
\label{eqmotsq}
\begin{split}
     &\dot{a} = - i \Delta_S a   - \frac{\kappa}{2} a - i \varepsilon_c a^\dg -i \LL( G_- b +G_+ b^\dg \RR) + \sqrt{\kappa_{Ei}} a_{in,i}(t)+ \sqrt{\kappa_{Eo}} a_{in,o}(t) + \sqrt{\kappa_I} a_{in,I}(t)\\
&\dot{a}^\dag = i \Delta_S a^\dg  - \frac{\kappa}{2}a^\dag +i \varepsilon_c^* a +  i \LL( G_-^* b^\dg +G_+^* b \RR)  + \sqrt{\kappa_{Ei}} a_{in,i}^\dg(t)+ \sqrt{\kappa_{Eo}} a_{in,o}^\dg(t) + \sqrt{\kappa_I} a_{in,I}^\dag(t) \\
     &\dot{b} = -i (\Delta_A   + G_2) b - \frac{\gamma_m}{2} b -i \varepsilon_m b^\dag-i G_+  a^\dg - i G_- ^* a     + \sqrt{\gamma_m} b_{in} \\  
          &\dot{b}^\dg = i (\Delta_A  + G_2) b^\dg- \frac{\gamma_m}{2} b^\dg +i \varepsilon_m^* b+ i G_+ ^* a + i G_- a^\dg    + \sqrt{\gamma_m} b_{in}^\dg \\  
\end{split}
\end{equation}

We solve the frequency-domain equations resulting from \eref{eqmotsq} numerically up to all orders. The response of each operator is written as arising from all the input fields, given as an example for the cavity operator as:
\begin{equation}
\label{afullsystemS21}
\begin{split}
a(\omega) & = m(\omega) a_{in,i}(\omega) +l(\omega) a^{\dg}_{in,i}(\omega) +m_o(\omega) a_{in,o}(\omega) +l_o(\omega) a^{\dg}_{in,o}(\omega) +\\
& + m_I (\omega)a_{in,I}(\omega) +l_I (\omega) a_{in,I} ^{\dg}(\omega)+ q(\omega) b_{in}(\omega) +r(\omega) b_{in}^\dg(\omega) \,.
     \end{split}
\end{equation}
One of the aims is to calculate the quadrature spectra of the mechanical resonator, in particular, the fluctuation energy as a function of the quadrature angle $\Theta$. The spectra of the mechanics is given as
\begin{equation}
\label{Sxtheta}
\begin{split}
      S_{x,\Theta}(\omega) & =  \langle (b^{\dg}(\omega)\exp (i \Theta)  + b (\omega)\exp (-i \Theta) ) (b^{\dg}(-\omega)\exp (i \Theta)  + b (-\omega)\exp (-i \Theta) ) \rangle =\\ 
     & =  \langle b^{\dg}(\omega) b^{\dg}(-\omega) \rangle \exp (2i \Theta) + \langle b^{\dg}(\omega) b(-\omega) \rangle + \langle b(\omega) b^{\dg}(-\omega) \rangle +  \langle b(\omega) b(-\omega) \rangle \exp (-2i \Theta)\,. \\
     \end{split}
\end{equation}
The correlators are calculated using \eref{afullsystemS21}, \eref{incorrelators}.


The quadrature variance, directly giving the energy of the quadrature in units of the zero-point energy is given as 
\begin{equation}
\begin{split}
     & \Delta X_{\Theta}^2 =  \frac{1}{2\pi} \int^{\infty}_{-\infty}  2 S_{x,\Theta}(\omega) d\omega \,.
          \end{split}
\end{equation}
In these units, the total energy of the mode is
\begin{equation}
\begin{split}
n_m = \frac{1}{4} \LL( \Delta X_{\Theta}^2 +  \Delta X_{\Theta+\pi/2}^2 \RR) \,.
          \end{split}
\end{equation}
For example, in the ground state this gives $n_m = \puoli$.

We use the standard nomenclature that the minimum of $ \Delta X_{\Theta}^2$ with respect to $\Theta$ is called the $X_1$ quadrature, and the maximum is $X_2$. The corresponding variances are labeled with the shorthand notations $ \Delta X_1^2$ and $ \Delta X_2^2$, respectively.

The final measured quantity is the output field of the field leaking out from the output coupler of the cavity, given as
\begin{equation}
\label{aout}
\begin{split}
a_{out,o} = \sqrt{\kappa_{Eo}} a - a_{in,o} \,,
          \end{split}
\end{equation}
which can be expressed as a sum of inputs the same way as in \eref{afullsystemS21}. For the first two terms, for example, this reads $a_{out,o}(\omega)  = m_{out,o}(\omega) a_{in,i}(\omega) +l_{out,o}(\omega) a^{\dg}_{in,i}(\omega)$. The spectral density of the corresponding voltage fluctuations is labeled $S_V(\omega)$. This is the quantity presented in the figures, given in units of quanta. The quadrature spectrum $S_{V,\Theta}(\omega)$ of the voltage fluctuations is given as in \eref{Sxtheta} by replacing the mechanics operators with those of the output field (\eref{aout}). For notational convenience, we use for the spectra of the minimum energy and maximum energy quadratures the notations $S_1(\omega)$ and $S_2(\omega)$, respectively. The final deductions regarding squeezing in the mechanical resonator are based on direct comparison of the measured phase-sensitive output spectrum to the predictions resulting from the presented formalism. In principle it is possible to obtain analytical results, but they are too complicated to be written down, but nevertheless, allow for numerical analysis.

Since theoretical expectations arise from nontrivial analysis, it is integral to make sure the calculations are sound. We verified that the results of the theoretical formalism are similar to earlier work in cases where comparison is possible. More specifically, the amount of squeezing our formalism gives is the same as in Ref.~15, as well as the total output spectrum reduces to their result in the limit of small internal cavity losses. If considering only the regular sideband cooling case, analytical results are available in Ref.~28, and our analysis agrees with those results.

For certain calibrations (section \ref{sec:detune}) we use the direct microwave transmission measurement through the cavity. The usual microwave transmission coefficient is given by
\begin{equation}
\label{S21}
\begin{split}
|S_{21}(\omega)| = \sqrt{m_{out,o}^2(\omega) +l_{out,o}^2(\omega) } \,.
          \end{split}
\end{equation}


\section{Calibrations}

The calibrated parameters of the experiment are listed in Table \ref{tab:calibpars}. In main text it was described how  $\gamma_m$ and the Bogoliubov mode ratio  $1/\tanh r = G_-/G_+$ were obtained. In this section, we discuss further details, and how the remaining parameters are calibrated.

\begin{table}[ht]
\caption{\emph{List of calibrated parameters}.}
\begin{center}
\begin{tabular}{c|cl|}
 symbol &  & explanation   \\
    \hline
  $G_-$  & & effective optomechanical coupling due to the red-detuned pump \\
  $\mathcal{G}$ & & gain of the system (amplifiers plus cabling) following the output port of the cavity \\
 $1/\tanh r$ & &  Bogoliubov mode ratio \\
 $\gamma_m$ & & intrinsic damping rate of the mechanical resonator \\
 $\kappa_{Ei}$ & & external damping rate of the cavity through the input port \\
 $\kappa_{Eo}$ & &  external damping rate of the cavity through the output port \\
 $\kappa_I$ & & internal damping rate of the cavity \\
 $\Delta_S$ & & symmetric detuning of the pump frequencies \\
 $\Delta_A$ & & asymmetric detuning of the pump frequencies  \\
  \hline
  \label{tab:calibpars}
\end{tabular}
\end{center}
\end{table}
%


\subsection{Effective coupling, system gain: $G_-$, $\mathcal{G}$}

\label{sec:calT}

Let us discuss the basis of the thermal calibration as shown in Fig.~2a in main text. The output spectrum $S_{V}(\omega)$ divided by system gain $\mathcal{G}$, when the pump is applied at the red sideband, supposing $\gamma_{\m{eff}} \ll \kappa$, is
\begin{equation}
\label{eq:cool}
\frac{S_V}{\mathcal{G}} = \frac{4 \kappa_{Ei} \kappa_I}{\kappa^2} n_I  + \frac{\kappa_{Ei}}{\kappa}\frac{\gamma_{\m{opt}} \gamma_{\m{eff}}}{\LL(\omega-\omega_m\RR)^2 + \gamma_{\m{eff}}^2/4} \LL( n_m^T - 2 \frac{\kappa_I}{\kappa}n_I \RR) \;,
\end{equation}
which is a Lorentzian centered at the cavity frequency. Here, the sideband cooling enhances damping by the amount
\begin{equation} 
\label{eq:gopt}
\gamma_{\m{opt}}= \frac{4 G_-^2 }{\kappa} \;,
\end{equation} 
so that the total (''effective'') damping of mechanics is
\begin{equation} 
\label{eq:geff}
\gamma_{\m{eff}}= \gamma_{\m{opt}} + \gamma_m \;.
\end{equation} 
Here, the effective coupling depends on the photon number induced by the pump:
\begin{equation} 
\label{eq:gminus}
G_- = g_0 \sqrt{n_-} \;.
\end{equation} 
%
We now suppose $\gamma_{\m{opt}} \ll \gamma_m$ such that cavity back-action is negligible.  Also $n_I \ll n_m^T$, such that we can neglect $n_I$ in the second term in \eref{eq:cool}. The first term involving $n_I$ would cause a shift in the base level. In any case, $n_I$ is essentially zero at the low pump powers. We fit a Lorentzian to each peak, obtaining the amplitude and linewidth for each temperature.

From \eref{eq:cool} we obtain a prediction for the peak area
\begin{equation} 
\label{eq:apeak}
A_{\m{peak}} = 2 \pi \mathcal{G} \frac{\kappa_{Ei}}{\kappa^2} 4 G_-^2  n_m^T = 2 \pi \mathcal{G} \frac{\kappa_{Ei}}{\kappa^2} 4 G_-^2 k_B T/\hbar \omega_m  \equiv \mathcal{K} T 
\end{equation} 
In cryogenic microwave measurements, $\mathcal{G}$ is poorly known and difficult to measure because of uncertainties in the attenuations of cables. For the same reason, the value of $G_-$ at a given input power to the cryostat is difficult to tell. In this calibration, therefore, we simply combine all (thus far) badly known parameters into a single coefficient $\mathcal{K}$. We fit $A_{\m{peak}}$ to a linear temperature dependence, obtaining Fig.~2a in the main text. Finally the equilibrium phonon number is obtained as $n_m^T  = A_{\m{peak}} k_B / (\hbar \omega_m \mathcal{K})$ for each data point.


For calibrating $G_-$ versus generator power, we study the peak width as a function of power (\eref{eq:geff}). According to \eref{eq:gopt}, \eref{eq:geff}, \eref{eq:gminus}, the peak width is proportional to the cavity photon number which depends linearly on power. One has to bear in mind that the width of the peak in the output spectrum, \eref{eq:cool}, directly gives $\gamma_{\m{eff}}$ only outside the strong coupling regime. The result is shown in \fref{Gredcalib}. We took into account the effect of around 6 \% change of $\kappa$ versus power.

\begin{figure}[htp]
 \includegraphics[width=7cm]{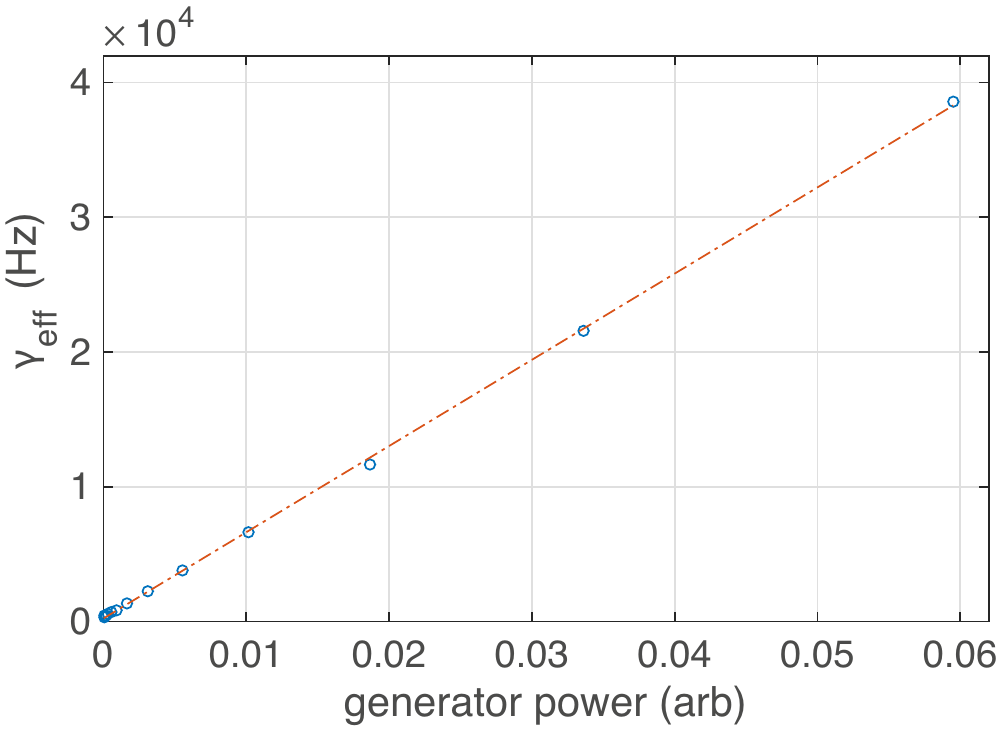}
  \caption{\emph{Calibration of effective coupling}. The largest power shown corresponds to an effective coupling $G_m/2\pi \simeq 79$ kHz.}
  \label{Gredcalib}
\end{figure}

For calibrating $ \mathcal{G}$, we use the basic sideband cooling data where example curves shown in black in Fig.~2b-e in main text. Each curve (about 20 curves at different power) are simultaneously fitted to theory with the same $ \mathcal{G}$. The theory is basically a more accurate version of \eref{eq:cool}, but allowing for the strong-coupling regime (peaks non-Lorentzian). We use the $G_-$ and $n_m^T$ just calibrated. The latter means that we know that at low pump powers, the mechanics has a given bath temperature. Hence in the fit, we enforce this, but otherwise allow for free $n_m^T$ and $n_I$ (note that these $n_m^T$ and $n_I$ are not those of the final squeezing experiment).

\subsection{Mechanical linewidth: $\gamma_m$}

In \fref{SummaryOfTemperatures} we show how the baths, to which either the cavity and the mechanics are coupled, heat up as a function the pump power(s). This heating ultimately limits the squeezing. The bath temperatures are not totally monotonous with respect to increasing pump power. We also sometimes observe the bath temperatures to slightly fluctuate during a cooldown, presumably due to changes in the configuration of microscopic defects in the sample, varying the thermal coupling from the pump into the bath. 


\begin{figure*}[htp]
 \includegraphics[width=0.8\linewidth]{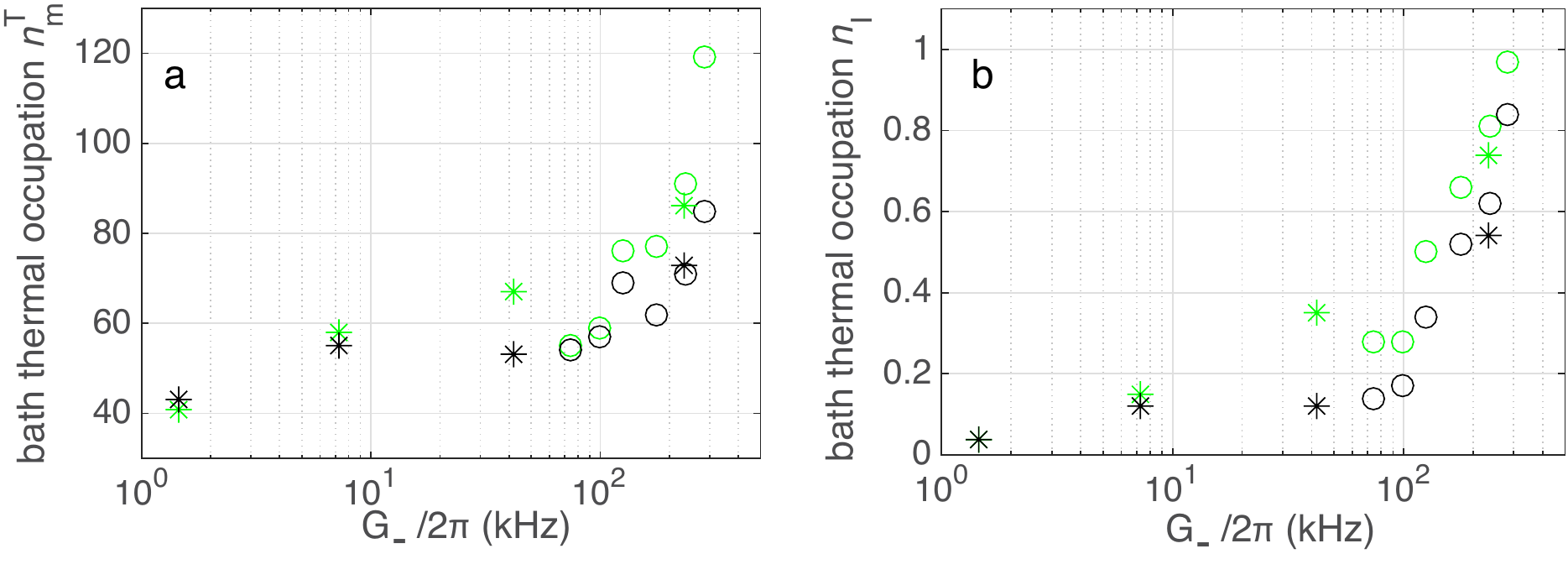}
  \caption{\emph{Bath temperatures as a function of effective coupling}. The data points are from the data shown in the main text in Fig.~2 (asterisk) and Fig.~3 (circles). As elsewhere, green refers to the BG mode, and black is the sideband cooling. (a), mechanics. (b), cavity.}
  \label{SummaryOfTemperatures}
\end{figure*}

Because the mechanical bath heats up appreciably at the highest pump powers, studying the mechanical linewidth $\gamma_m$ at the base temperature of the cryostat (at which temperature the measurements are done) may not be sufficient. In \fref{gammam_vs_T} we show the dependence of the mechanical linewidth as a function of the refrigerator temperature. As seen in the figure, the mechanical damping does not depend on temperature more than the scatter of the data points in the region $\lesssim 60$ mK. This temperature range covers the bath temperatures at all pump powers. Hence we can use the same value for $\gamma_m$ over the whole pump power range.

\begin{figure*}[ht]
 \includegraphics[width=0.4\linewidth]{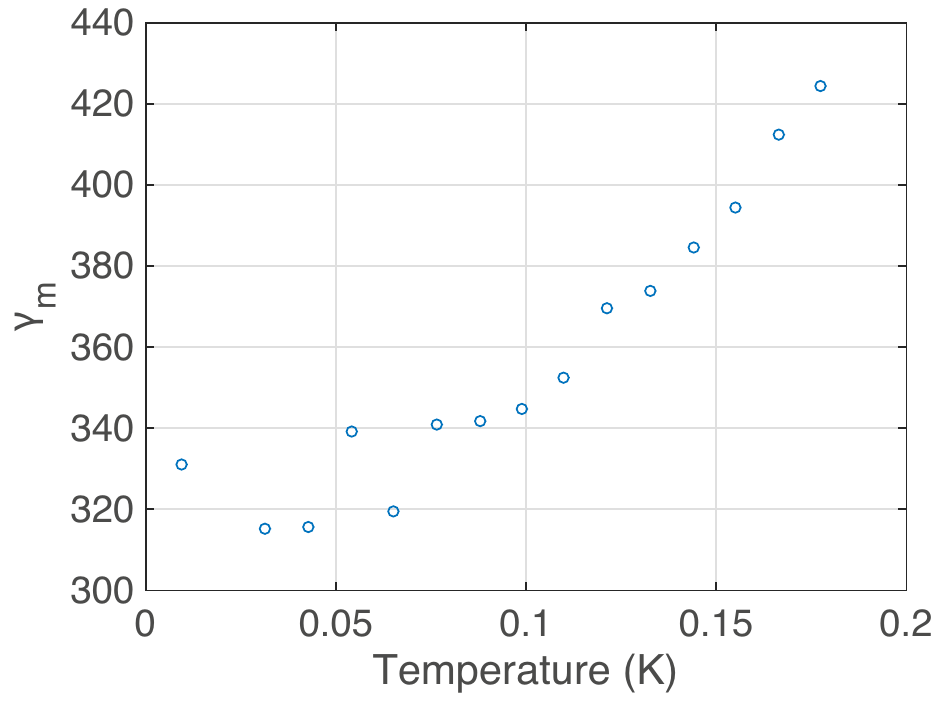}
  \caption{\emph{Mechanical intrinsic linewidth $\gamma_m$ vs cryostat temperature}.}
  \label{gammam_vs_T}
\end{figure*}

\subsection{Cavity linewidth: $\kappa_{{Ei}} $, $\kappa_{{Eo}} $, $\kappa_{{I}} $ }

The input $\kappa_{{Ei}} $ and output $\kappa_{{Eo}} $ coupling rates are determined by an electromagnetic simulation in the actual device structure. Based on typical spread in the dimensions of fabricated devices, we estimate around 10 \% error margins for them. The total linewidth $\kappa = \kappa_{{Ei}} + \kappa_{{Eo}} + \kappa_I$ is determined from a fit to the measured transmission in case optomechanical effects are not relevant. A subtraction gives the internal decay rate $\kappa_I$.

\subsection{Thermalization of the Bogoliubov mode}


In section \ref{sec:calT} we discussed the thermalization of the bare mechanical mode. Because the BG mode follows the similar Hamiltonian, \eref{eq:bghamilt}, a similar temperature dependence holds for the energy of the BG mode as well. We calculate the green theory curve in Fig.~2a in main text as follows. From the output spectrum based on \eref{Sxtheta}, \eref{aout} we numerically calculate predictions for the areas under the sideband peaks separately for both the BG and mechanical modes. The green theory curve is by the ratio of these two, a factor of 1.42, above the bare mechanics theory curve.


\subsection{Effect of pump detuning: $\Delta_S$, $\Delta_A$}

\label{sec:detune}

Unless we are operating at the exact sideband co-resonance $\Delta_\pm = \pm \omega_m $, the resonance condition is lost and physics becomes more complicated than simply sideband cooling of the BG mode. The scale at which the "cavity" detuning $\Delta_S$ should be correct is $\kappa$. For the "mechanics" detuning $\Delta_A$, the situation is not so clear, but we estimate the scale is between $\gamma_m$ and $\gamma_{\m{opt}}$, and hence one has to be more cautious. We could fit the detuning to the spectra, but we calibrate it independently using the cavity linear response to a probe signal ($S_{21}$ microwave transmission), see \fref{detune}.

It is relatively simple to accurately set the asymmetric detuning, \eref{eq:bghamilt}, to zero because we can measure the mechanical frequency accurately, and possible changes are expected to be minor, and on the other hand, the cavity transmission is sensitive to a non-zero value. Setting the symmetric detuning, however, is more challenging because the cavity frequency has some dependence on pump power, and hence we have to first calibrate the cavity frequency as a function of pump power.

An example of the detuning calibration under the double-pump condition is given in \fref{detune}. For the fit, we ignore the parametric effects which have a negligible effect on the transmission. The fit gives $\Delta_S \simeq 11$ kHz $\pm 400$ Hz, $\Delta_A \simeq 510$ Hz $\pm 120$ Hz, although the resonance condition was nominally set. Notice that the transmission is quite sensitive to both detunings which allows for small error bars.

At this level of detuning, the affect on squeezing is negligible. For the data in Fig.~2b-e in the main text no careful calibration of the symmetric detuning $\Delta_S$ was done, but it is estimated to be around 40 kHz based on the shift of the BG mode. Even this amount does not cause appreciable degradation of squeezing.

\begin{figure*}[htp]
 \includegraphics[width=0.5\linewidth]{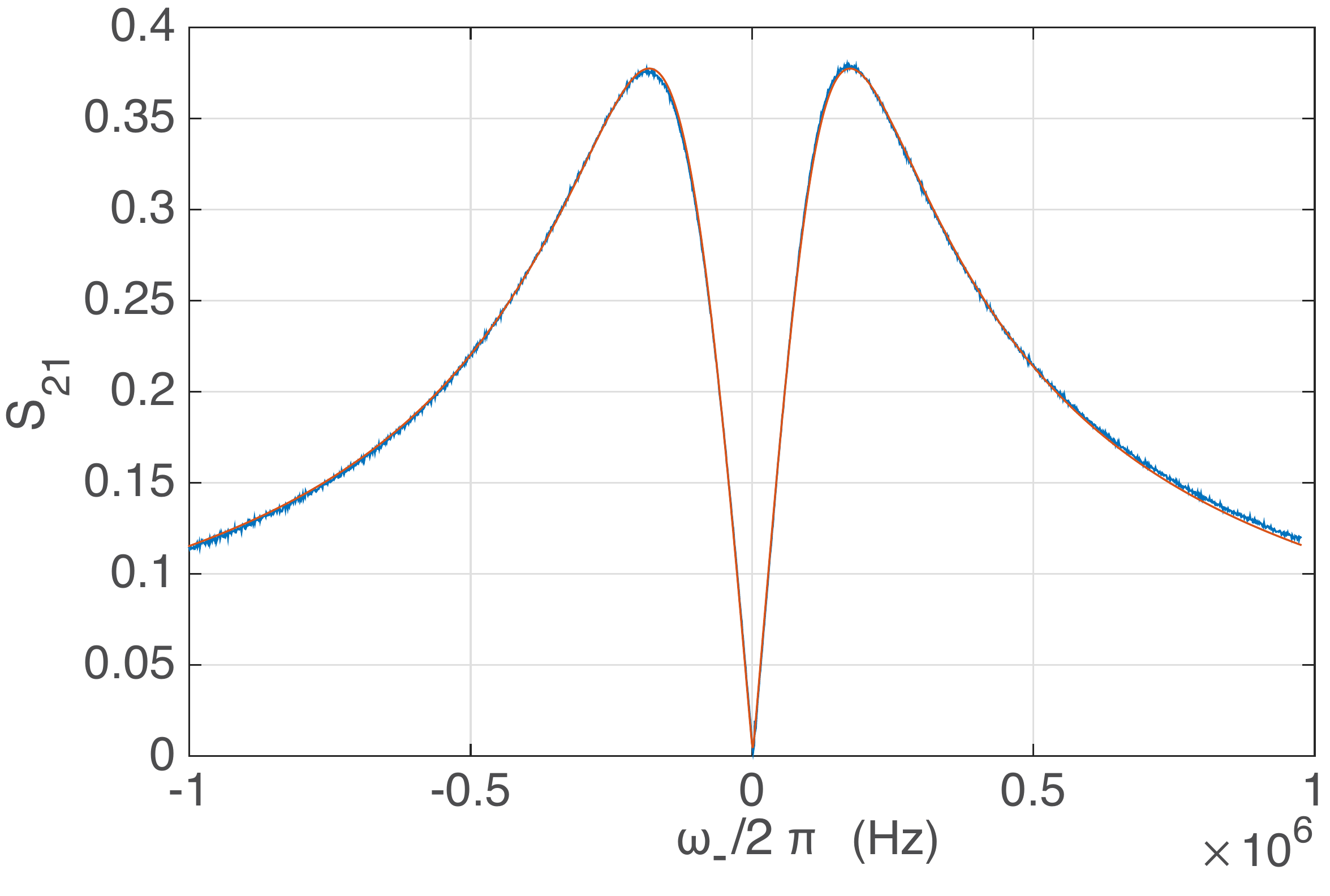}
  \caption{\emph{Calibration of pump detuning}. Plotted is the measured cavity transmission corresponding to Fig.~3d. The theory fit is superimposed on the data.}
  \label{detune}
\end{figure*}

\subsection{Parametric modulation}

\label{sec:paramcalib}


One can make an independent measurement of the cavity parametric effect by the use of a large detuning; $|\Delta_\pm| \gg \kappa$ (\eref{eq:deltapm}) but such that the center frequency stays at $\omega_c$. Expressed with the "cavity" and "mechanics" detunings $\Delta_S$ and $\Delta_A$ in \eref{eq:bghamilt}, we have $\Delta_S \sim 0$ and $\Delta_A \gg \kappa$. With the high detuning, optomechanical effects which take place within the scale of $\kappa$ about the sideband resonance, become negligible.

We carried out such a measurement in the scheme approximately the same as Fig.~3d in main text. We used $\Delta_\pm \simeq \omega_m \pm (2\pi )Ê\cdot 4$ MHz. Although we used the same generator power as in Fig.~3d, the effective couplings are smaller because of higher detuning from the cavity; we obtain $G_- /2\pi \simeq 180$ kHz.

The theory curves in   \fref{paramcalib} are calculated for a parametrically modulated oscillator (no optomechanics). We obtain a good agreement by using the values $|\varepsilon_c|/2\pi = 42$ kHz and the phase $\arg \LL( \varepsilon_c \RR) = -\pi/2$. The magnitude is in reasonable agreement to what we have for the squeezing spectra with a similar $G_- $ in Fig.~3c. With different values of $\arg \LL( \varepsilon_c \RR) $, the pattern would shift by a corresponding amount with respect to LO phase. In particular, the Kerr effect of the cavity corresponds to $\arg \LL( \varepsilon_c \RR) = 0$ which is excluded.

\begin{figure*}[htp]
 \includegraphics[width=0.9\linewidth]{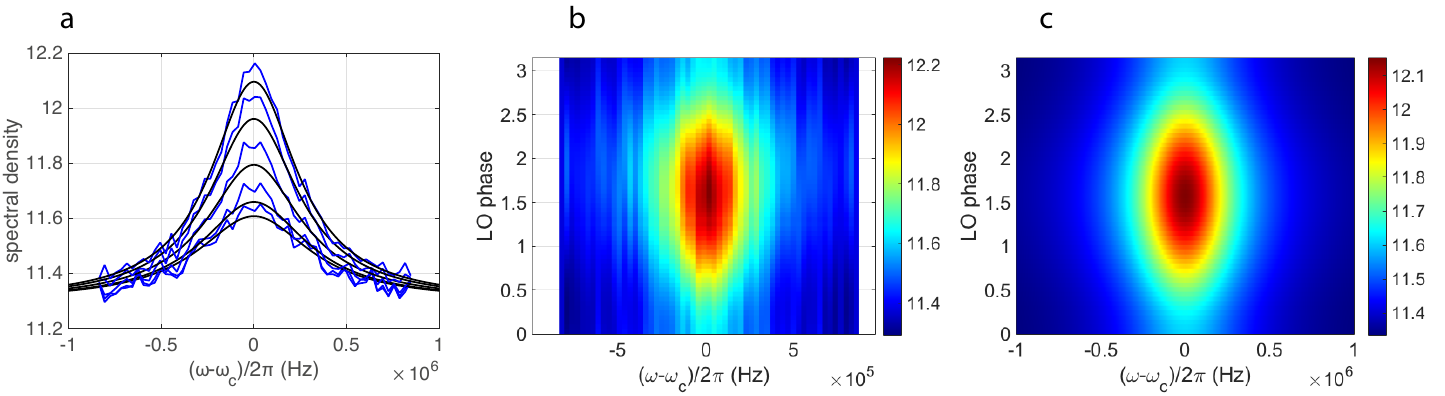}
  \caption{\emph{Calibration of parametric effects}. (a) Curves plotted at different LO phase values at $\pi/10$ steps from 0 to $\pi/2$, from bottom to top. The theory fit is superimposed on the data. (b) The same data plotted as surface. (c) Corresponding theory plot.}
  \label{paramcalib}
\end{figure*}

\subsubsection{Different types of parametric modulation}

\label{sec:diffparam}

Here we discuss how various types of parametric modulations would show up in the quadrature spectra. As seen in   \fref{paramplay}, changes in phase or amplitude, or modulation to the mechanics instead of cavity, each cause characteristic features in the spectra, each of which are inconsistent with data. As seen in (a) and (b), the experiment and theory clearly show a pattern symmetric with respect to the LO phase $\pi/2$. In particular, the case  \fref{paramplay}c corresponds to the cavity Kerr effect, causing very asymmetric pattern. In (d), the modulation of the mechanical frequency tends to sharpen the peaks. In either case, the effect on squeezing is minor.

\begin{figure*}[ht]
 \includegraphics[width=0.6\linewidth]{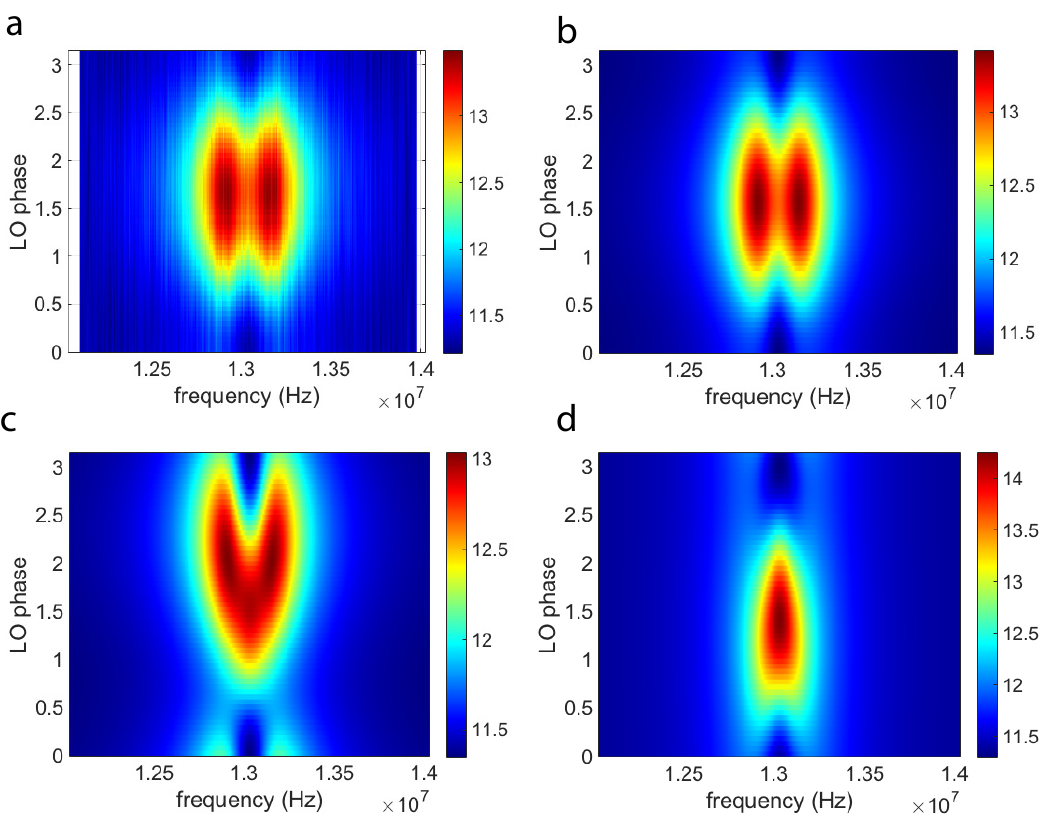}
  \caption{\emph{Squeezing spectra with different parametric effects}. The setup is as in Fig.~3d in main text. (a) Experiment. (b), Theory as in Fig.~3d. (c) Same as (b), but with $\arg \LL( \varepsilon_c \RR) = 0 $. (c), Parametric modulation to mechanics: $\varepsilon_c = 0$, $\varepsilon_m = 4.2 $ kHz.}
  \label{paramplay}
\end{figure*}




As mentioned in the main text, it is possible to obtain a good fit of the Bogoliubov mode data (that is, the total output spectrum) without imposing any parametric modulation. However, one finds out that far too little phase dependence would be predicted for the quadrature spectra. We illustrate this in \fref{noparamp}. The color codes are as in Fig.~3d in main text. The parameters are $n_m^T = 91$, $n_I = 1.07$, $\Delta X_1^2= 0.79$. One has to use an elevated cavity internal temperature to explain the enhanced emission of BG mode. In the real case, the parametric modulation acts to boost up the emission. The squeezing is only little affected, nevertheless.

\begin{figure*}[htp]
 \includegraphics[width=0.5\linewidth]{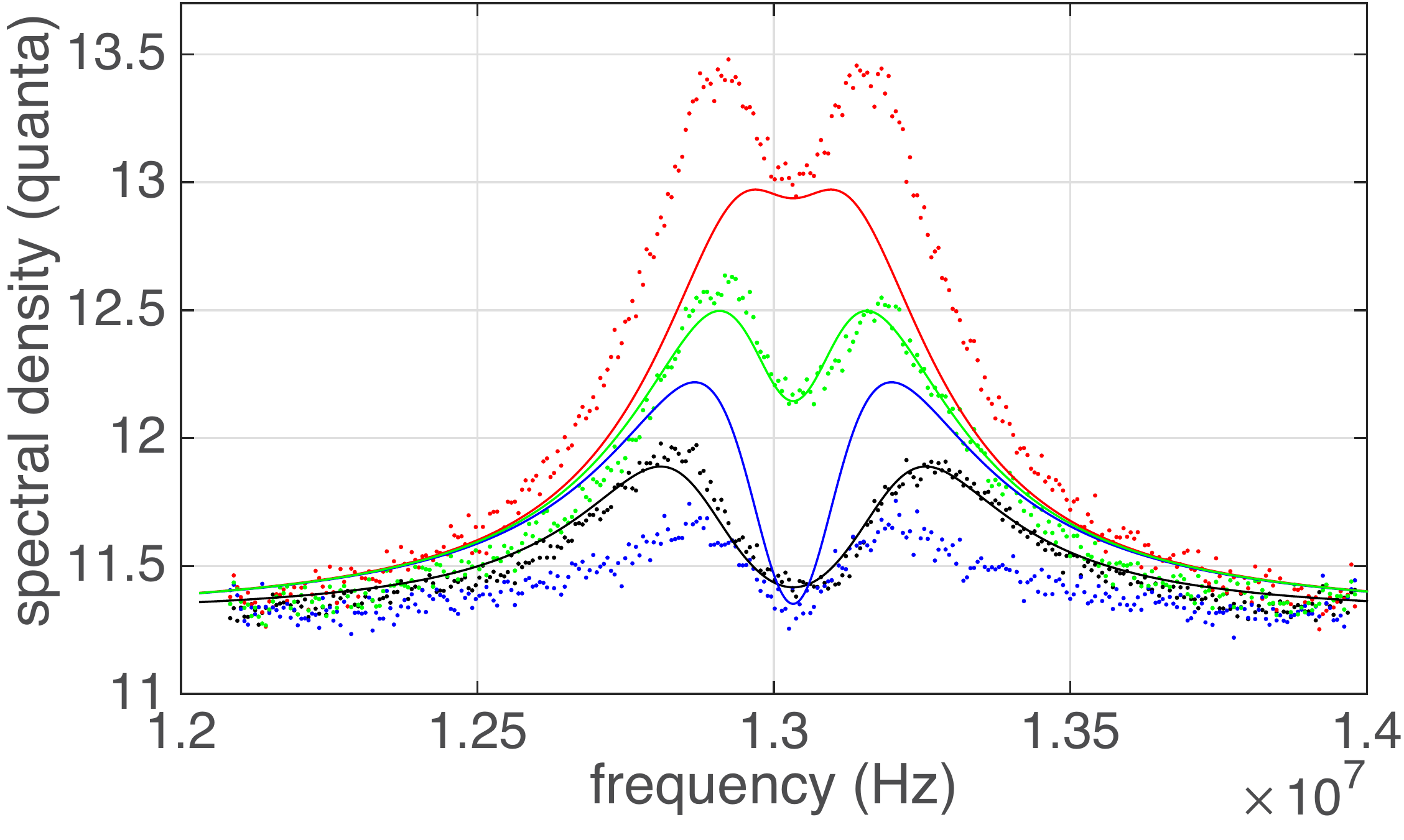}
  \caption{\emph{Modeling without parametric modulation}. As Fig.~3d in main text, but with $\varepsilon_c = 0$.}
  \label{noparamp}
\end{figure*}

\section{Error analysis}

Next we discuss the error analysis of the final results demonstrating squeezing. We will notice the imprecision is dominated by uncertainty of the system gain at the output side. The other significant contributions to the imprecision come from those of the mechanics bath thermalization, mechanical intrinsic linewidth, and of the BG mode ratio.


A general remark is that the shapes of the BG mode and quadrature spectra are sensitive to most parameters, the bath temperatures of both the mechanics and the cavity, $n_m^T$ and $n_I$ in particular. The third parameter on which squeezing significantly depends on and whose independent determination has non-negligible error margins, is the mechanical intrinsic linewidth $\gamma_m$.

We make error estimate for the squeezed quadrature variance using standard error propagation as
\begin{equation}
\begin{split}
\delta \LL( \Delta X_1^2\RR) = \sqrt{\sum_i  \LL( \delta y_i \frac{\partial \Delta X_1^2}{\partial y_i} \RR)^2}\,,
          \end{split}
\end{equation}
where the statistical as well as systematic (calibration) uncertainties of variables $y_i$ are marked by $\delta y_i$. The derivatives $\frac{\partial \Delta X_1^2}{\partial y_i}$ are obtained numerically. 


The error terms are summarized in Table \ref{tab:error}. If possible, we list both statistical and systematic error for each parameter. The quoted statistical errors for the three adjustable parameters  $n_m^T$,  $n_I$,  $\varepsilon_c$ are from direct fits to the squeezing spectra. For some calibrated parameters, there can be also systematic uncertainties on top of those obtained from the fit in question. The statistical errors are quoted as $2\sigma$ error bars. The uncertainty for each parameter is obtained as the largest of the statistical error and systematic error.



The uncertainty $\delta n_m^T$ of the starting temperature of the mechanics bath is particularly important to consider. At low pump powers, an error in the bath temperature calibration would show up as a scaling factor in the Lorentzian peak, easily going unnoticed. The real situation, however, is better because towards the strong-coupling regime, the starting temperature has a more complicated influence than just scaling the peaks. The linear fit to \eref{eq:apeak} (plotted in Fig.~2a) gives a negligible uncertainty, only $\sim 1$ \%. A systematic error estimate is based on the fact that the data points at $T < 20$ mK, which region marks the base temperature, show more scatter than those at higher temperature, probably marking fluctuations in thermalization. We get another estimate for $\delta n_m^T$ from the scatter. The standard error of the mean gives $\delta n_m^T \simeq 3$, or approximately 10 \% uncertainty. One more way to obtain $\delta n_m^T$ is from the direct statistical error of the fits to the quadratures, yielding $\delta n_m^T \simeq 1.2$.


The uncertainty of the fitted value of $\gamma_m$ in the data shown in Fig.~1e in main text is  $\delta \gamma_m/2\pi \simeq 14$ Hz. This value is considered as the systematic uncertainty. Because $\gamma_m$ is not an adjustable parameter in the end (when fitting to the quadrature spectra), it does not have a statistical uncertainty because of the way we carry out the analysis. Nevertheless, one might ask the question of what if $\gamma_m$ would have a statistical uncertainty. For the latter purpose, we can make a test where in the fits we in fact consider it as an adjustable parameter. Some of the true adjustable parameters were fixed during this test for numerical stability. The direct statistical error this way gives an estimate $\delta \gamma_m/2\pi \simeq 5$ Hz, which is negligible in comparison to the systematic uncertainty.


The uncertainties of $n_I$ and $\varepsilon_c$ are obtained as a direct statistical error of the fits to the quadratures.

The uncertainty of the cavity linewidth $\kappa$ is obtained from Lorentzian fits to the measured cavity transmission. The statistical error from a given fit is negligible, but we obtain some scatter between different repetitions over a longer time period. Standard deviation of the latter gives $\delta \kappa/2\pi \sim 10$ kHz. Based on typical spread in the dimensions of fabricated devices, we estimate around 10 \% error margins for $\kappa_{{Ei}}$ and $\kappa_{{Eo}}$, and hence that for $\kappa_I$ is the sum of all three. We note the squeezing is very insensitive to how $\kappa$ is distributed between different channels.

The confidence level for $G_-$ when fitted to \eref{eq:geff} is $\pm 2$ \%.

\begin{table}[ht]
\caption{\emph{Error budget}. The numbers are for the optimum squeezing, i.e., Fig.~3d in main text, $G_- = $ 235 kHz. The last column tells the contribution of the error term to the total uncertainty of the squeezing. The frequencies are given in linear frequencies (Hz).}
\begin{center}
\begin{tabular}{c|c|c|c|c|c|c|}
 parameter $ y_i$ &  value & statistical $\delta y_i$ & systematic $\delta y_i$ &statistical  $\delta y_i \frac{\partial \Delta X_1^2}{\partial y_i}$ & systematic $\delta y_i \frac{\partial \Delta X_1^2}{\partial y_i}$ & percentage  \\
    \hline
 $n_m^T$ & 91 &  1.2 & 3 & 0.006 & 0.015 & 11  \\
 $n_I$ & 0.81 & 0.02  & - & 0.005 & - & 4\\
 $\varepsilon_c$ & 56 kHz & 1.2 kHz & - & 0.001 &- & 1 \\
 $G_-$ & 235 & - & 5 kHz & - & 0.01 & 7 \\
  $\mathcal{G}$ & arb & 2 \% & 11 \% & 0.01  & 0.06 &41 \\
 $1/\tanh r$ & 1.43 &- & 0.1 & - & 0.014 & 10 \\
 $\gamma_m$ & 330 Hz & 5 Hz & 14 Hz & 0.006 &  0.016 & 11 \\
 $\kappa_{Ei}$ & 50 kHz & - & 5 kHz & - & $ 10^{-3}$ & 2 \\
 $\kappa_{Eo}$ & 270 kHz & - & 30 kHz & -& 0.0039  & 3 \\
 $\kappa_I$ & 330 kHz & - & 40 kHz & - & 0.016 & 11 \\
 $\Delta_S$ & $11$ kHz & $400$ Hz & -& $ 10^{-5}$ & - & 0 \\
 $\Delta_A$ &  $510$ Hz & $ 120$ Hz & - & $10^{-5}$ & - & 0 \\
  \hline
  \label{tab:error}
\end{tabular}
\end{center}
\end{table}

For $ \mathcal{G}$, the direct statistical error is small, but we also have to propagate errors from the parameters $G_-$ and $n_m^T $ which are used to obtain $ \mathcal{G}$. The error due to $G_-$ is negligible, but that due to $n_m^T $ equals the confidence level of $n_m^T $, around 10 \%. In order to obtain the corresponding error in squeezing, we enforce a change in $ \mathcal{G}$ from the calibrated value and re-run the fits. As seen in Table \ref{tab:error}, the gain error dominates the squeezing uncertainty.

Uncertainty for the pump power ratio $1/\tanh r = G_- / G_+$ is obtained as systematic uncertainty of the amount 0.1 equaling the drift mentioned in the main text.


Combining the terms in Table \ref{tab:error} we obtain
\begin{equation}
\begin{split}
\delta \LL( \Delta X_1^2\RR) \simeq 0.08 \,,
          \end{split}
\end{equation}
which is the final uncertainty estimate.

For smaller pump powers, the uncertainties are larger because the peak in the cold quadrature is lower and signal-to-noise hence smaller and statistical uncertainties begin to dominate.



We can also demonstrate that the theory curves are rather sensitive to the parameters. In \fref{badfits} we show plots as in Fig.~3 d in main text, but one of the important parameters is intentionally offset by 20 \% from the nominal value to the direction which degrades squeezing, while other parameters are kept unchanged. In each case, the agreement is clearly worsened, while squeezing is affected around 10 \%. 


\begin{figure*}[htp]
 \includegraphics[width=0.95\linewidth]{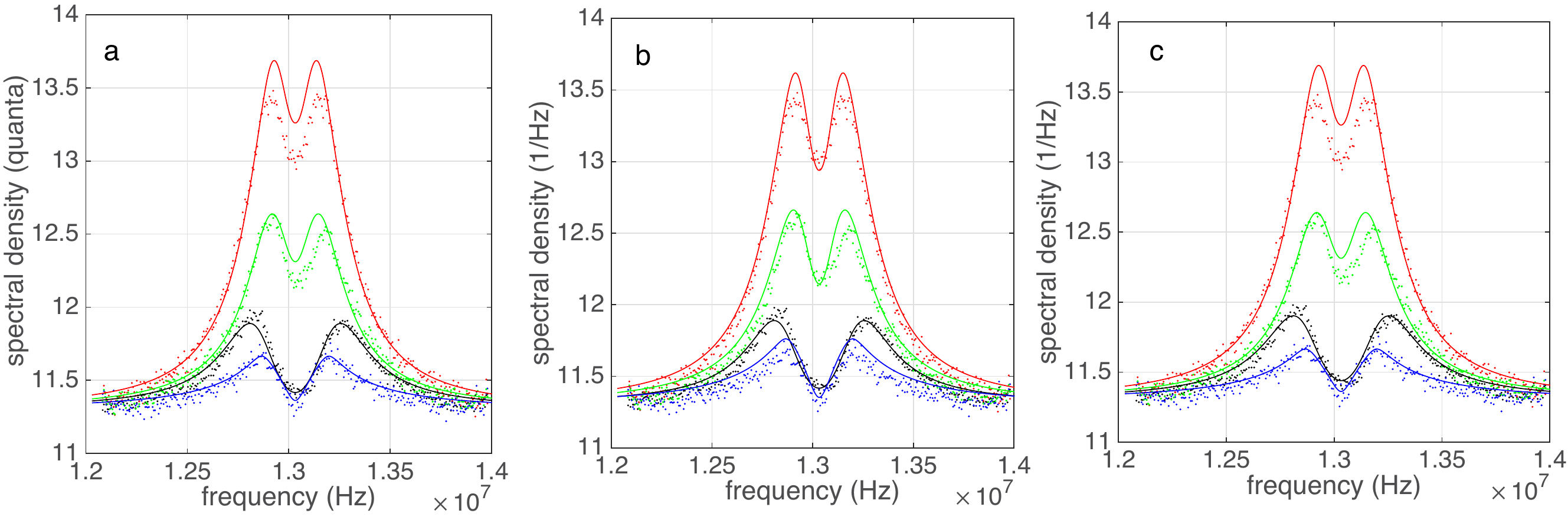}
  \caption{\emph{Tests of sensitivity to parameters}. (a), $n_m^T = 109$, (b), $n_I = 0.97$, (c), $\gamma_m/2\pi = 396$ Hz.}
  \label{badfits}
\end{figure*}

\vspace{1cm}




 \end{widetext}


\end{document}